\documentclass[useAMS,usenatbib]{mn2e}

\usepackage{graphicx}
\usepackage{amsmath}
\usepackage{amssymb}
\usepackage{color}
\usepackage{subfig}
\usepackage[breaklinks,colorlinks,citecolor=blue,linkcolor=red]{hyperref} 
\usepackage[all]{hypcap}

\newcommand\msun{\, \rm M_\odot}

\newcommand\kms{\, \rm km\,s^{-1}}

\newcommand\eout{{e_{\rm out}}}
\newcommand\einin{{e_{12}}}
\newcommand\einout{{e_{3}}}

\newcommand\aout{{a_{\rm out}}}
\newcommand\ainin{{a_{12}}}
\newcommand\ainout{{a_{3}}}
\newcommand\amax{{a_{\rm 3, max}}}
\newcommand\mmbh{{M_{\rm SMBH}}}

\title[BH and NS mergers in GN: the role of triples]{Black hole and neutron star mergers in Galactic Nuclei: the role of triples}
\author[G. Fragione et al.]{  \parbox{\textwidth}{Giacomo Fragione$^{1}$\thanks{E-mail: giacomo.fragione@mail.huji.ac.il}, Nathan W. C. Leigh$^{2,3,4}$, Rosalba Perna$^{4,5}$ \vspace*{0.3cm}}\\
$^1$Racah Institute for Physics, The Hebrew University, Jerusalem 91904, Israel\\
$^2$Departamento de Astronom\'ia, Facultad de Ciencias F\'isicas y Matem\'aticas, Universidad de Concepci\'on, Concepci\'on, Chile\\
$^3$Department of Astrophysics, American Museum of Natural History, Central Park West and 79th Street, New York, NY 10024 \\
$^4$Department of Physics and Astronomy, Stony Brook University, Stony Brook, NY 11794-3800, USA\\
$^5$Center for Computational Astrophysics, Flatiron Institute,  New York, NY 10010, USA}

\hypersetup{draft}

\begin{document}

\maketitle

\begin{abstract}
Nuclear star clusters that surround supermassive black holes (SMBHs) in galactic nuclei are thought to contain large numbers of black holes (BHs) and neutron stars (NSs), a fraction of which form binaries and could merge by Kozai-Lidov oscillations (KL). Triple compact objects are likely to be present, given what is known about the multiplicity of massive stars, whose life ends either as a NS or a BH. In this paper, we present a new possible scenario for merging BHs and NSs in galactic nuclei. We study the evolution of a triple black hole (BH) or neutron star (NS) system orbiting an SMBH in a galactic nucleus by means of direct high-precision $N$-body simulations, including Post-Newtonian terms. We find that the four-body dynamical interactions can increase the KL angle window for mergers compared to the binary case and make BH and NS binaries merge on shorter timescales. We show that the merger fraction can be up to $\sim 5$--$8$ times higher for triples than for binaries. Therefore, even if the triple fraction is only $\sim 10\%$--$20\%$ of the binary fraction, they could contribute to the merger events observed by LIGO/VIRGO in comparable numbers.
\end{abstract}


\section{Introduction}

The LIGO-Virgo collaboration has recently released a catalogue of compact object (CO) mergers due to gravitational wave (GW) emission, comprised of ten merging black hole (BH-BH) binaries and one merging neutron star (NS-NS) binary \citep{ligo2018}. Studying the possible mechanisms that lead to BH and NS mergers is currently an active area of research.  Several scenarios for their origins have been proposed. The possibilities include isolated binary evolution through a common envelope phase \citep{bel16b}, chemically homogeneous evolution in short-period stellar binaries \citep{mand16,march16}, stellar triples \citep{ant17,sil17}, quadruple systems \citep{fragk2019,liu2019} and Kozai-Lidov (KL) mergers of binaries in galactic nuclei \citep{antoper12,hamer18,hoan18,fragrish2018,grish18}. These mergers could either occur in active galactic nuclei (AGN) disks \citep[e.g.][]{mckernan18,secunda18} or in star clusters \citep{askar17,baner18,frak18,rod18}. While these models typically predict roughly the same rates ($\sim\ \mathrm{few}$ Gpc$^{-3}$ yr$^{-1}$), their respective contributions could potentially be disentagled using the observed eccentricity, spin, mass and redshift distributions \citep[see e.g.][]{olea16,samas18,zevin18,Perna2019}.

The wide literature on dynamically-induced mergers has focused on GW signals generated from merging BH-BH binaries in open and globular clusters. However, galactic nuclei typically have total stellar masses comparable to the combined stellar masses of open cluster systems, and a significant fraction of globular clusters in a given galaxy \citep{boker2004,cote2006}. Thus, galactic nuclei represent a promising environment where CO binaries can form dynamically, then harden and finally merge due to the emission of GW radiation \citep*{olea09,ant16,petr17,fra2019,rass2019}.

Compared to open and globular clusters, studies focused on galactic nuclei have generally either concentrated on how large-scale nuclear star cluster dynamics ($\gtrsim 1$ pc) can drive CO binaries to merge within a Hubble time \citep{ant16,petr17,gondan2018}, or on how these binaries merge due to eccentric KL oscillations imparted from the enormous SMBH potential \citep[$\lesssim 1$ pc;][]{antoper12,hoan18,fragrish2018,hamer18}.  

Triple stars are likely not rare in open clusters and possibly even globular clusters, in particular for massive stars whose lives end either as a NS or a BH \citep{duq91,ragh10,sa2013AA,tok14a,tok14b,duns2015,sana2017}. Little attention has been devoted to the role of merging binary COs in triple systems in galactic nuclei, where they can orbit a central SMBH. Previous studies have primarily focused on determining the merger rates from isolated triple BHs \citep{ant17,sil17} or triple BHs in globular clusters \citep{antcha2016}. Compared to binary systems, triples have an additional degree of freedom and KL oscillations can be imparted to the inner binary either by the SMBH or by the third companion \citep{lid62,koz62}. The interplay between these three KL mechanisms working together can give rise to a rich and complicated dynamical evolution for the inner binary \citep{grishlai2018}. This could potentially lead to a larger number of mergers, thtat can be further enhanced by possible resonances \citep{hamerslai2017}.

\begin{figure} 
\centering
\includegraphics[scale=0.45]{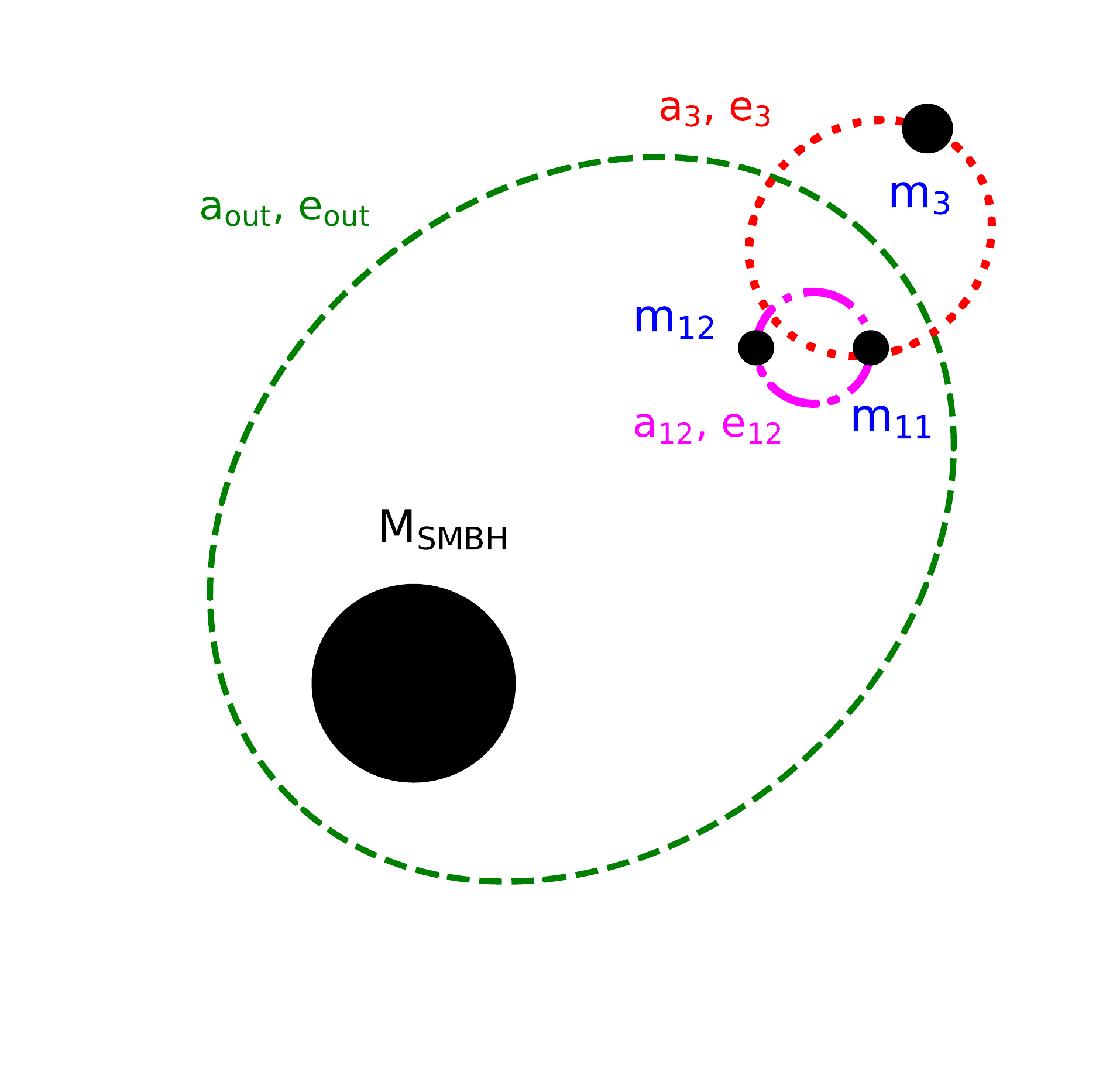}
\caption{The four-body system studied in the present work. We denote the mass of the SMBH as $\mmbh$, the masses of the components of the inner binary (in the BH/NS triple) as $m_{11}$ and $m_{12}$, and the mass of the third companion (in the BH/NS triple) as $m_3$. The semi-major axis and eccentricity of the outer orbit are denoted by $\aout$ and $\eout$, respectively. For the BH/NS triple, we denote with $\ainin$ and $\einin$ the semi-major axis and eccentricity of the inner orbit, and $\ainout$ and $\einout$ the semi-major axis and eccentricity of the outer orbit.}
\label{fig:fourbody}
\end{figure}

In this paper, we propose a new scenario for merging BHs and NSs in galactic nuclei. For the first time, we study the evolution of a triple composed of COs orbiting an SMBH in a galactic nucleus and compare it to the CO binary case \citep{fragrish2018}. We consider a four-body system consisting of a triple BH-BH-BH/NS-NS-NS and an SMBH (see Figure~\ref{fig:fourbody} and Table~\ref{tab:quant}). We make a systematic and statistical study of these systems by means of direct high-precision $N$-body simulations, including Post-Newtonian (PN) terms up to order PN2.5. Compared to secular evolution, our approach allows us to correctly follow the large eccentricity values attained by CO binaries via KL cycles \citep{antognini14}.
 
The paper is organized as follows. In Section~\ref{sect:bintrip}, we discuss what is to date known about the multiplicity properties of BHs and NSs in galactic nuclei and star clusters. In Section~\ref{sect:kozait}, we discuss the relevant timescales for the systems considered in this paper. In Section~\ref{sect:bhnsmergers}, we present our numerical methods to determine the rate of BH and NS mergers in triples orbiting massive black holes, for which we discuss the results. Finally, in Section~\ref{sect:conc}, we discuss the implications of our findings for future observations of mergers involving BHs and NSs, and draw our conclusions.

\section{Multiplicity}
\label{sect:bintrip}

In this section we briefly review what is known from the literature about multiplicity (i.e., binaries and triples) in the Galactic field, open clusters, globular clusters and nuclear star clusters.

\subsection{The Galactic field}
\label{field}

\citet{ragh10} updated the seminal work of \citet{duq91} by selecting from the Hipparcos catalogue a volume-limited sample of field solar-type primary stars in the solar neighbourhood.  The authors find that the observed fractions of objects that are single, double, triple and higher-order systems are, respectively, 56 $\pm$ 2, 33 $\pm$ 2, 8 $\pm$ 1 and 3 $\pm$ 1 per cent. Subsequent work showed that most triples composed of low-mass stars tend to be approximately co-planar, whereas the inner and outer orbits of triples composed of massive stars tend to be more inclined relative to each other \citep[e.g.][]{moe18}.

\begin{table*}
\caption{Description of important quantities used in the text.}
\centering
\begin{tabular}{lc}
\hline
Symbol & Description\\
\hline\hline
$\mmbh$ & Mass of the SMBH \\
$m_{11}$ & Mass of the first component of the inner binary in the BH/NS triple \\
$m_{12}$ & Mass of the second component of the inner binary in the BH/NS triple \\
$m_3$ & Mass of the third companion in the BH/NS triple \\
$\aout$ & Semi-major axis of the orbit of the triple around the SMBH \\
$\eout$ & Eccentricity of the orbit of the triple around the SMBH \\
$\ainin$ & Semi-major axis of the inner orbit in the BH/NS triple \\
$\einin$ & Eccentricity of the inner orbit in the BH/NS triple \\
$\ainout$ & Semi-major axis of the outer orbit in the BH/NS triple \\
$\einout$ & Eccentricity of the outer orbit in the BH/NS triple \\
\hline
\end{tabular}
\label{tab:quant}
\end{table*}

\subsection{Open clusters}
\label{OCs}

The young star-forming region Taurus-Auriga has an exceedingly low density, such that dynamical interactions involving multiples should be very rare, if they occur at all.  In fact, Taurus-Auriga is the only known open cluster with a triple fraction higher than that observed in the Galactic field \citep{ragh10,leigh13}.  \citet{kraus11} performed a high-resolution imaging study to characterize the multiple-star populations in Taurus-Auriga. They found that between $\sim$ $2/3$--$3/4$ of all targets are multiples composed of at least two stars. Thus, only $\sim$ $1/4$--$1/3$ of their objects are single stars.

More nearby open clusters (OCs) are also known to have comparably high multiplicity fractions (see \citet{leigh13} for a more detailed review of the following studies).  The Hyades \citep{patience98}, Pleiades \citep{mermilliod92,bouvier97} and Praesepe \citep{mermilliod99,bouvier01} have binary fractions of, respectively, 35\%, 34\% and 40\%, and triple fractions of, respectively, 6\%, 3\% and 6\%.  

Finally, old open clusters, with masses and densities typically higher than the above OCs, have also been observed to harbour high binary fractions.  For example, for the old ($\sim$ 4 Gyr) OC M67, \citet{fan96} observed a binary fraction of 45\%.  For comparison, the old ($\sim$ 7 Gyr) OC NGC 188 has an observed multiplicity fraction of 27\% \citep{geller09}.  Following \citet{leigh13}, if we adopt the observed value for the ratio between the fractions of binaries and triples from \citet{latham96}, or $f_{\rm t}$/$f_{\rm b}$ $\sim$ 0.1, we infer triple fractions for M67 and NGC 188 of, respectively, 5\% and 2\%.

\subsection{Globular clusters} \label{GCs}

In globular clusters (GCs), the situation is quite different.  Due to their much higher central densities compared to OCs, the multiple star populations in massive GCs can most efficiently be studied using photometry.  The pioneering study of \citet{milone12} analyzed the main-sequence (MS) binary populations in a sample of 59 GCs. The authors find photometric binary fractions ranging from less than a percent to a few tens of percent, and a previously reported \citep{sollima07} anticorrelation between the binary fraction and the total cluster mass. \citet{sollima08} argued that this trend can arise assuming an universal initial binary fraction combined with the dynamical disruption of binaries by (mostly) interloping single stars. Here, the disruption of soft binaries in the cluster core generates the observed anticorrelation, combined with the evaporation of single stars from the cluster outskirts which serves to increase the binary fraction throughout the cluster in preferentially low-mass clusters evaporating at the fastest rates \citep[e.g.][]{fregeau2009}.

Other than identifying binaries photometrically above and/or to the red of the main sequence in the cluster colour-–magnitude diagram \citep[e.g.][]{milone12}, binaries can also be found at higher energies as exotic objects like low-mass X-ray binaries \citep[LMXBs; e.g.][]{hut91}, millisecond pulsars \citep[MSPs; e.g.][]{verbunt87} and cataclysmic variables \citep[CVs; e.g.][]{pooley06,cohn10}. Blue straggler (BS) formation is also thought to involve binary or triple stars \citep[e.g.][]{perets09,knigge09,leigh11}.  This can occur in several different ways, such as mass transfer within a binary, collisions during encounters involving binaries or even some triple-based mechanism \citep[e.g.][]{leigh11,geller13} such as KL-induced mergers \citep[e.g.][]{perets09,naoz14}.

To date, only one triple star system is known to exist in the dense environments characteristic of GCs \citep[e.g.][]{prodan12}. The system in question, called 4U 1820-30, lives near the centre of the GC NGC 6624.  It consists of a low-mass X-ray binary with a NS primary and a WD secondary, in orbit with a period $\sim$ $685$ s.  There is also a large luminosity variation for this system (a factor $\sim$ 2) with a period of $\sim$ 171 days.  This longer period is thought to be due to the presence of a tertiary companion \citep{grindlay88}.

\subsection{Nuclear clusters} \label{NCs}

Much less is known of the multiple star populations in nuclear star clusters (NSCs).  Binaries can only be detected at home, in the Milky Way's Galactic nucleus.  But here, significant reddening intervenes along the relevant line of sight, rendering observations of the Galactic Centre challenging at most wavelengths.  If present, however, they should experience direct dynamical interactions with single stars in such dense environments on short timescales, and most likely be disrupted in the process \citep[e.g.][]{leigh16,leigh18}.

There are at least 3 known stellar binaries within the central $0.2$ pc of the Galactic Centre  \citep[e.g.][]{naoz2018,ott1999,rafelski2007}.  What's more, it is still possible that many of the observed S-stars are unresolved binaries \citep[e.g.][]{naoz2018}.  Even S0-2, which has the closest pericentre passage to Sgr A* and has received relatively careful scrutiny in recent years, could still be an unresolved binary \citep{chu2018}.

More recently, \citet{hailey18} discovered $13$ X-ray sources within $\sim 1$ pc from Sgr A*, at least in projection. The authors argue that the accretors must be BHs instead of NSs for the majority of their sample. This argument is based on attempting to relate the properties of the observed spectra to those of other NSs known to be present in the Galactic Centre.  These authors argue that the new X-ray sources show, in this regard, unique spectral features relative to the other known NSs in the Galactic Centre. 

As for triples, to the best of our knowledge, there are none yet known to exist in our Galactic Centre. If they do exist, they likely formed from a relatively recent episode of star formation, which there is plenty of observational evidence to support \citep{bart09}. This is because in the high-density, high-velocity dispersion environment of the Galactic Centre, triples should be rapidly destroyed due to dynamical interactions with single stars \citep[e.g.][]{leigh18}.

\section{Timescales}
\label{sect:kozait}

A triple system made up of an inner binary that is orbited by an outer companion undergoes KL oscillations in eccentricity whenever the initial mutual orbital inclination of the inner and outer orbit is in the window $i_0\sim 40^\circ$--$140^\circ$. The KL oscillations occur on a secular quadrupole timescale \citep{antognini15,nao16}
\begin{equation}
T_{\rm KL}=\frac{8}{15\pi}\frac{m_{\rm tot}}{m_{\rm 3b}}\frac{P_{\rm 3b}^2}{P_{\rm bin}}\left(1-e_{\rm 3b}^2\right)^{3/2}\ ,
\end{equation}
where $m_{\rm 3b}$ is the mass of the outer body and $e_{\rm 3b}$ its eccentricity, $m_{\rm tot}$ is the total mass of the triple system, and $P_{{\rm bin}}\propto a_{{\rm bin}}^{3/2}$ and $P_{{\rm 3b}}\propto a_{{\rm 3b}}^{3/2}$ are the orbital periods of the inner and outer binary, respectively. At the quadruple order of approximation (inner test particle and outer circular orbit), the maximal eccentricity is a function of the initial mutual inclination
\begin{equation}
e_{in,max}=\sqrt{1-\frac{5}{3}\cos^2 i_0}\ .
\label{eqn:emax}
\end{equation}
We note that there is also a dependence on the relative orbital orientation \citep{kino1999}. At this order of approximation, the inner binary eccentricity approaches unity when $i_0$ approaches $90^\circ$ (if not quenched by other sources of precession). In the case the outer orbit is eccentric, the inner eccentricity can reach almost unity even if the initial inclination is outside of the $i_0\sim 40^\circ$-$140^\circ$ KL range \citep[octupole order of approximation;][]{naozf13a}. This happens over the octupole timescale
\begin{equation}
T_{\rm oct}=\frac{1}{\epsilon}T_{\rm LK}\ ,
\label{eqn:tlkoct}
\end{equation}
where the octupole parameter is defined as 
\begin{equation}\label{oc1}
\epsilon={m_{\rm bin,1}-m_{\rm bin,2}\over m_{\rm bin,1}+m_{\rm bin,2}}\frac{a_{\rm bin}}{a_{\rm 3b}}\frac{e_{\rm 3b}}{1-e_{\rm 3b}^2}\ ,
\end{equation}
with $m_{\rm bin,1}$ and $m_{\rm bin,2}$ being the masses of the objects in the inner binary\footnote{According to \citet{antognini15}, the octupole timescale is rather $T_{\rm oct}=T_{\rm LK}/\epsilon^{1/2}$.}. In the case of a CO binary, the large values reached by the eccentricity of the inner binary during the KL excursions make its merger time shorter since it efficiently dissipates energy when $e \sim e_{\rm max}$ \citep[e.g.][]{blaes2002,thomp2011,antoper12,antognini15}. Tidal bulges and relativistic precession can suppress KL cycles \citep[e.g.][]{nao16}. In particular, for COs the relevant process is general relativistic precession that operates on a timescale
\begin{equation}
T_{GR}=\frac{a_{\rm bin}^{5/2}c^2(1-e_{\rm bin}^2)}{3G^{3/2}(m_{\rm bin,1}+m_{\rm bin,2})^{3/2}}\ .
\end{equation}
Thus, the KL oscillations of the orbital elements are damped by relativistic effects in the region of parameter space where $T_{KL}>T_{GR}$. On the other hand, \citet{naoz2013} showed that there can also be resonant behaviour when the KL and 1PN (Post-Newtonian) timescales are comparable, which could lead to eccentricity excitations.

\begin{figure} 
\centering
\includegraphics[scale=0.5]{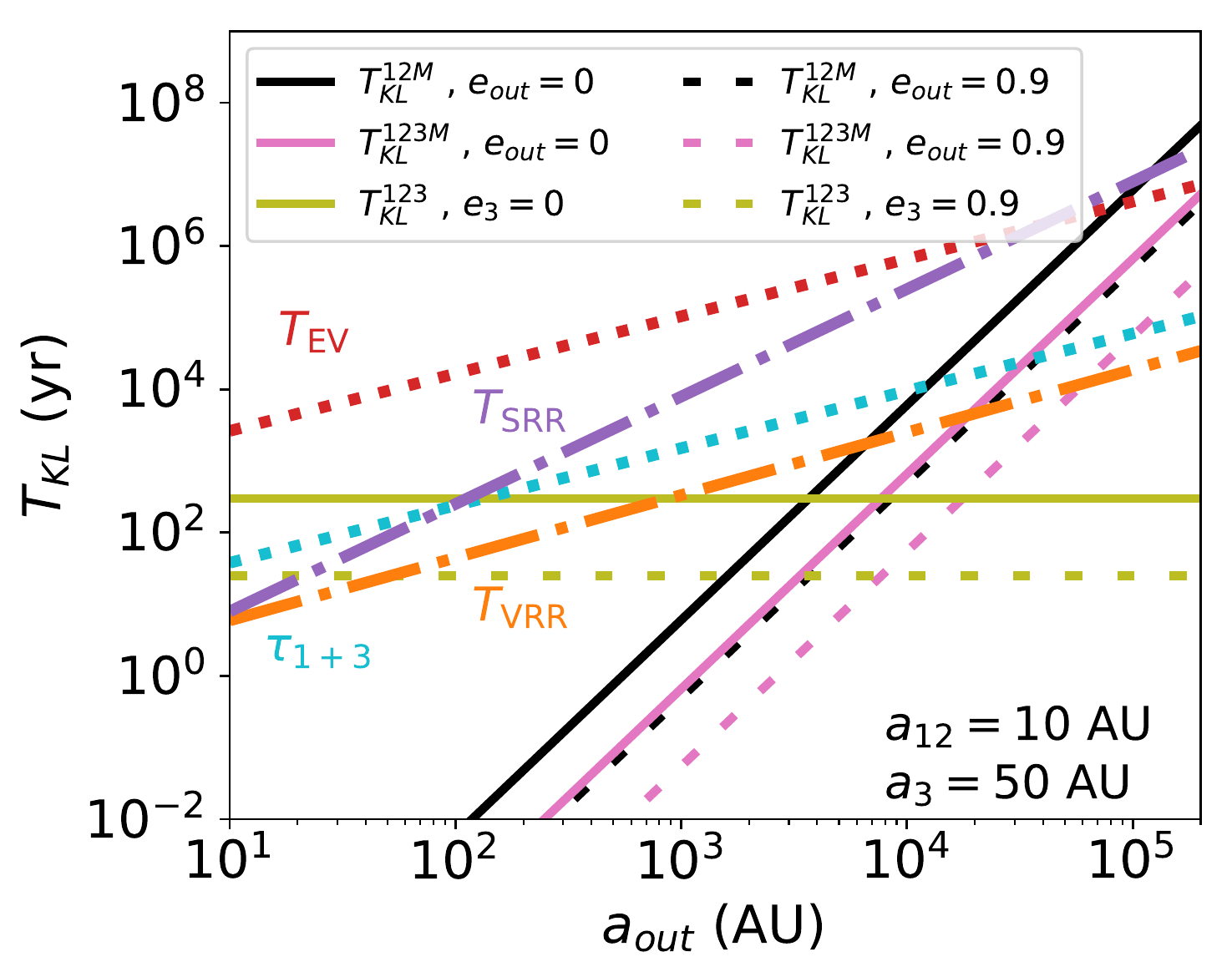}
\includegraphics[scale=0.5]{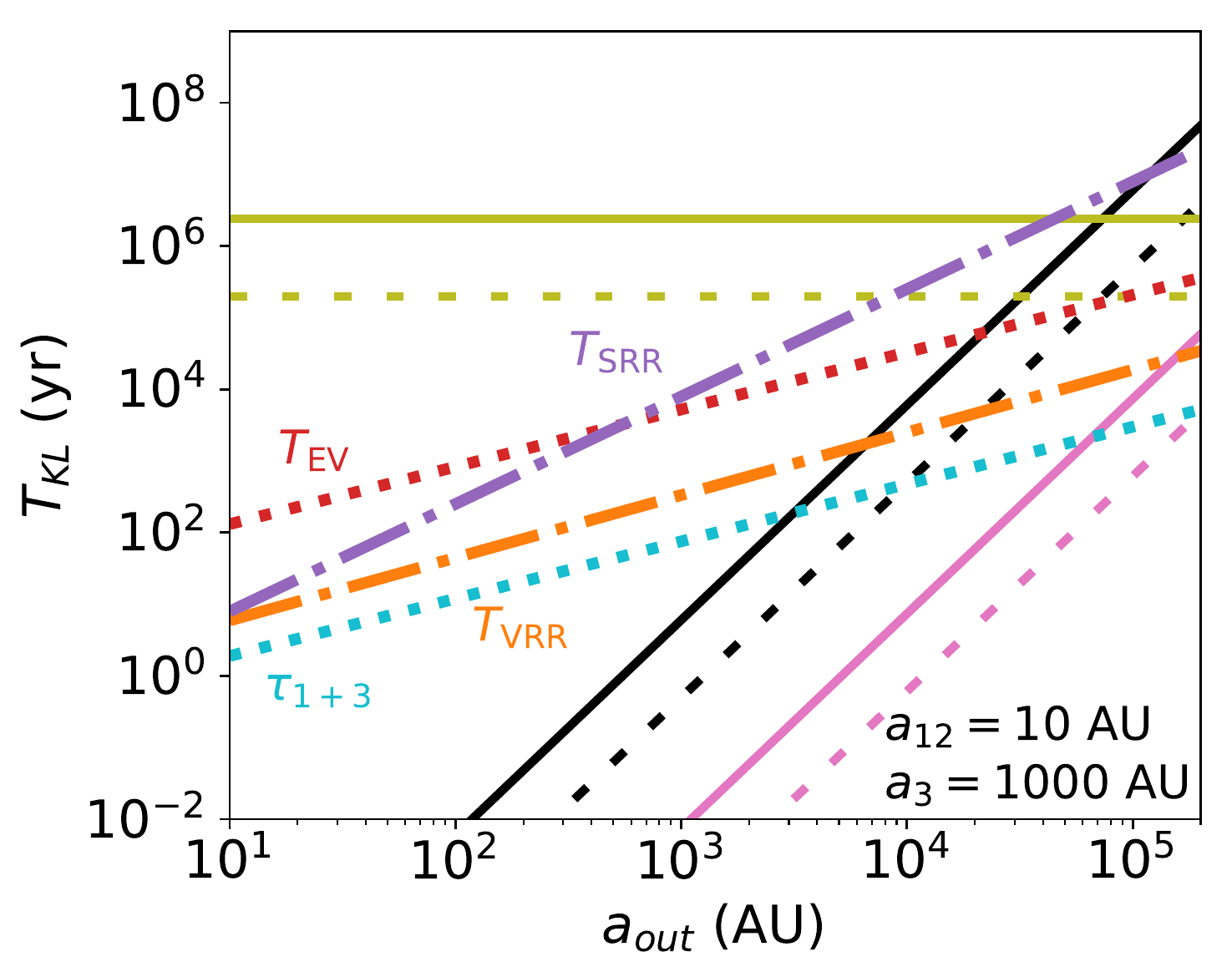}
\includegraphics[scale=0.5]{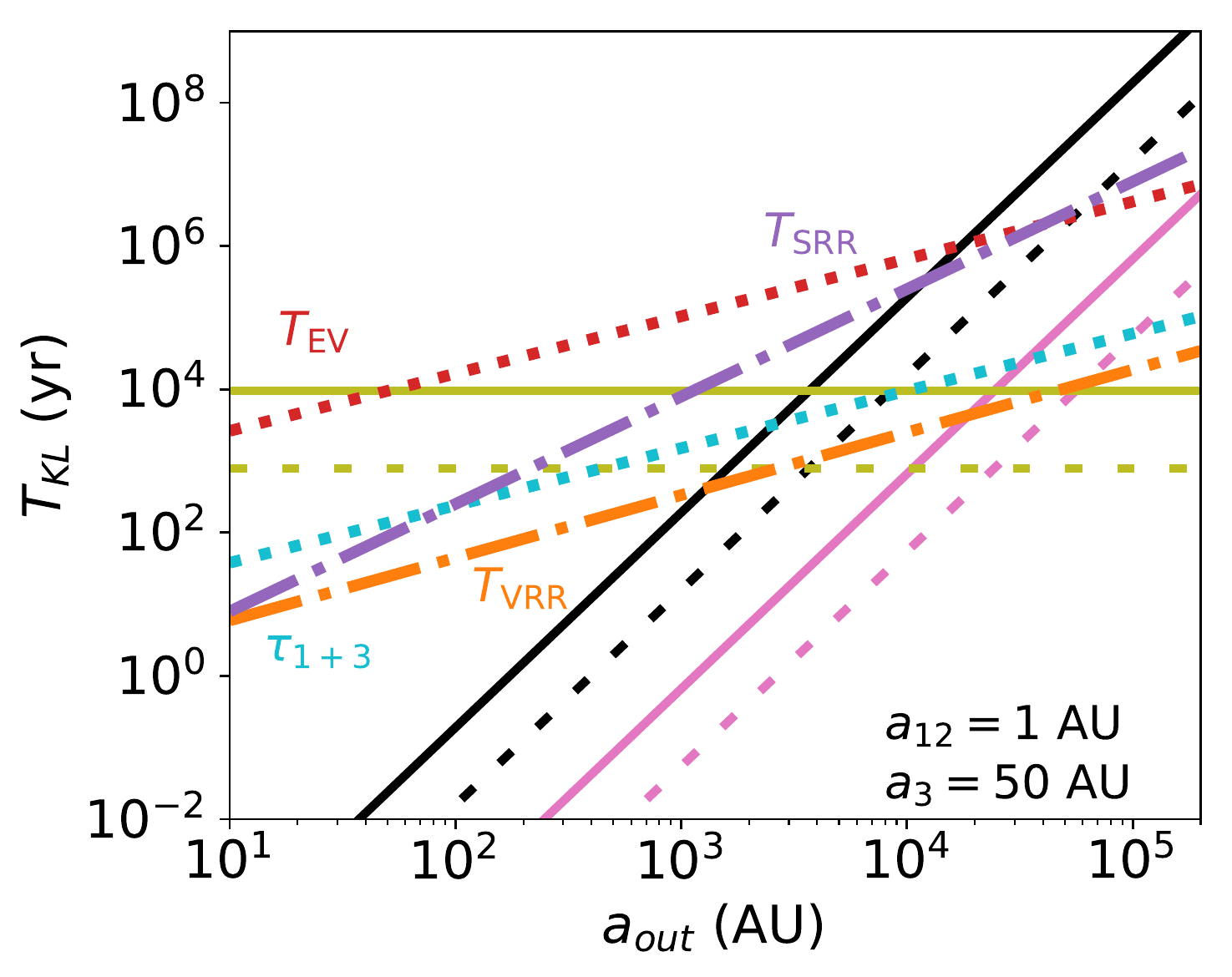}
\caption{KL timescales (Eqs.~\ref{eqn:tkl12m}-\ref{eqn:tkl123m}-\ref{eqn:tkl123}), 1+3 encounter timescale for triples (Eq.~\ref{eqn:tenc13}), and evaporation timescale (Eq.~\ref{eqn:evap}) as a function of distance from the SMBH for different parameters of the triple. The masses of the BHs are $m_{11}=m_{12}=m_{3}=10\msun$.}
\label{fig:KLtimesc}
\end{figure}

At the quadrupole order, the Hamiltonian representing the evolution of our systems can be described as  composed of three three-body Hamiltonians \citep{hamers2015}: (i) SMBH and the innermost binary of the CO triple; (ii) SMBH and the binary made up of the centre-of-mass of the inner binary and the third companion of the CO triple; (iii) third companion and the inner binary of the CO triple. Therefore, for the systems studied in this paper, there are three relevant (quadrupole) KL timescales. The first can be calculated by considering the SMBH as a perturber of the inner binary of the CO triple. In this case $m_{\rm tot}\approx \mmbh$ and
\begin{eqnarray}
T_{\rm KL}^{12M}&=&\frac{8}{15\pi}\frac{P_{\rm out}^2}{P_{\rm 12}}\left(1-e_{\rm out}^2\right)^{3/2}=\nonumber\\
&=&\frac{16}{15 G^{1/2}}\frac{a_{\rm out}^3}{\mmbh}\frac{(m_{11}+m_{12})^{1/2}}{a_{\rm 12}^{3/2}}\left(1-e_{\rm out}^2\right)^{3/2}\ .
\label{eqn:tkl12m}
\end{eqnarray}
The second KL timescale is obtained when considering the effect of the SMBH on the binary made up of the centre-of-mass of the inner binary and the third companion of the CO triple. Also in this case, $m_{\rm tot}\approx \mmbh$ and
\begin{eqnarray}
T_{\rm KL}^{123M}&=&\frac{8}{15\pi}\frac{P_{\rm out}^2}{P_{\rm 123}}\left(1-e_{\rm out}^2\right)^{3/2}\nonumber\\
&=&\frac{16}{15 G^{1/2}}\frac{a_{\rm out}^3}{\mmbh}\frac{m_\mathrm{t}^{1/2}}{a_{\rm 3}^{3/2}}\left(1-e_{\rm out}^2\right)^{3/2}\ ,
\label{eqn:tkl123m}
\end{eqnarray}
where $m_\mathrm{t}=m_{11}+m_{12}+m_3$. Of course, $T_{\rm KL}^{123M}<T_{\rm KL}^{12M}$. The last KL timescale refers to tidal perturbations of the third BH/NS on the inner binary of the CO triple ("internal" KL timescale of the triple)
\begin{eqnarray}
T_{\rm KL}^{123}&=&\frac{8}{15\pi}\frac{m_{11}+m_{12}+m_{3}}{m_{\rm 3}}\frac{P_{\rm 3}^2}{P_{\rm 12}}\left(1-e_3^2\right)^{3/2}\nonumber\\
&=&\frac{16}{15 G^{1/2}}\frac{m_\mathrm{t}}{m_{\rm 3}}\frac{a_{\rm 3}^3}{m_3}\frac{(m_{11}+m_{12})^{1/2}}{a_{12}^{3/2}}\left(1-e_{\rm 3}^2\right)^{3/2}\ .
\label{eqn:tkl123}
\end{eqnarray}

Figure~\ref{fig:KLtimesc} shows these three relevant KL timescales as a function of distance from the SMBH, for different parameters of the triple. For reference, we fix $m_{11}=m_{12}=m_{3}=10\msun$. The internal KL timescale for the CO triple is mainly set by the (outer) semi-major axis $a_3$. For the case $\ainin=10$ AU and $\ainout=50$ AU, $T_{\rm KL}^{123}$ is smaller than $T_{\rm KL}^{12M}$ and $T_{\rm KL}^{123M}$ when $\aout\gtrsim 10^3$ AU for $\mmbh=4\times 10^6\msun$ (i.e. a Milky Way-like galaxy). For the same $\ainin$ and $\ainout=1000$ AU, $T_{\rm KL}^{123}$ becomes the smallest only for $\aout\gtrsim 10^4$ AU--$10^5$ AU. On the other hand, if we reduce the inner semi-major axis of the triple to $\ainin=1$ AU, the typical distance at which the internal triple KL mechanism operates on the shortest timescale increases by a factor of $\sim 2$. 

In Figure~\ref{fig:KLtimesc}, we also plot two additional dynamical timescales relevant for triples in galactic nuclei. The first one is the typical encounter timescale for triples with other (single) stars \citep{leigh11},
\begin{eqnarray}
\tau_{1+3}&=& 0.78\ \mathrm{Myr}\ (1-f_{\rm b}-f_{\rm t})^{-1}\left(\frac{\sigma}{300\kms}\right)\times\nonumber\\
&\times& \left(\frac{1\msun}{m_*}\right)\left(\frac{1\ \mathrm{AU}}{a_3}\right)\left(\frac{10^6\msun\ \mathrm {pc}^{-3}}{\rho}\right)\ ,
\label{eqn:tenc13}
\end{eqnarray}
where $f_b$ and $f_t$ are the fractions of binaries and triples in the galactic nucleus, $\rho$ is the density in stars and $\sigma$ is the velocity dispersion, and $m_*$ the mass of a typical object in the system. For timescales $\gtrsim \tau_{1+3}$, the triple survives around the SMBH, but with orbital elements different from the initial ones. The second timescale is the evaporation timescale $T_{\rm EV}$ due to the subsequent dynamical interactions with field stars in the dense environment of a galactic nucleus. A triple can evaporate when
\begin{equation}
\frac{E_t}{m_\mathrm{t}\sigma^2}\lesssim 1\ ,
\end{equation}
where $E_t$ is the internal orbital energy of the triple and $m_\mathrm{t}=m_{11}+m_{12}+m_{3}$. This process happens on an evaporation timescale\footnote{To compute the relevant quantities in Eq.~\ref{eqn:tenc13} and Eq.~\ref{eqn:evap}, we have assumed the stars are distributed according to a \citet{bahcall76} cusp and normalization as given by Eq. 4 in \citet{hoan18}.} \citep{binntrem87}
\begin{eqnarray}
T_{\rm EV}&=& 180\ \mathrm{Myr} \left(\frac{m_\mathrm{t}}{30\msun}\right)\left(\frac{\sigma}{300\ \mathrm{km s}^{-1}}\right)\times\nonumber\\
&\times&\left(\frac{1\ \mathrm{AU}}{a_{\rm 3}}\right)\left(\frac{10^{6}\msun\ \mathrm{pc}^{-3}}{\rho}\right)\left(\frac{15}{\ln \Lambda}\right)\ ,
\label{eqn:evap}
\end{eqnarray}
where $\ln \Lambda$ is the Coulomb logarithm. Both the encounter timescale and the evaporation timescale are of the order of $\sim$ Myr for a Milky Way-like galaxy. For timescales $\sim T_{\rm EV}$, the triple is expected to be broken up by the many dynamical interactions with other stars.

Finally, there are two other relevant timescales, that we show in Figure~\ref{fig:KLtimesc}, which are the timescales associated with resonant relaxation (RR). This effect is the result of torques originating from correlated encounters with stars orbiting the SMBH. For a given star or CO, RR affects both the eccentricity (scalar RR; SRR) and the inclination (vector RR; VRR) of the orbital plane with respect to the SMBH on timescales \citep[e.g.][]{hamer18}
\begin{equation}
T_{\rm VRR}=\frac{\pi a_{\rm out}^{3/2}}{G^{1/2}}\frac{M_{\rm SMBH}^{1/2}}{m_*\sqrt{N_*(\aout)}}\ ,
\end{equation}
and
\begin{equation}
T_{\rm SRR}=\frac{\pi a_{\rm out}^{3/2}}{2G^{1/2}}\frac{M_{\rm SMBH}^{1/2}}{m_*}\ ,
\end{equation}
respectively, where $N_*$ is the number of stars orbiting the SMBH within $\aout$.

In the case of a CO binary orbiting an SMBH, we expect most of the mergers due to the KL mechanism to happen when $i_0\sim 90$ \citep{fragrish2018}. When a CO triple is present, the relative interplay among the three different KL mechanisms can bring the CO orbits into the relevant KL window, and hence induce mergers via GW emission because of a faster energy dissipation rate supplied by the increasing eccentricity. This is also favoured by the possible resonance between nodal precession and KL oscillations, whenever these two mechanisms operate on comparable timescales \citep{hamerslai2017,grishlai2018}.

\section{N-Body Simulations}
\label{sect:bhnsmergers}

In this section we quantify the physical properties of the merging triples with direct N-body simulations in order to compute their merger rates and assess whether they are enhanced with respect to the binary case.

\begin{table}
\caption{Models: triple type, slope of the BH mass function ($\beta$), slope of the distribution of $\aout$ ($\alpha$), distribution of $\einin$ and $\einout$ ($f(e)$), maximum outer semi-major axis of the triple ($\amax$), merger fraction from the $N$-body simulations ($f_{\rm merge}$). For comparison, the second part of the table summarises the results of \citet{fragrish2018}.}
\centering
\begin{tabular}{cccccc}
\hline
Triple Type & $\beta$ & $\alpha$ & $f(e)$ & $\amax$ (AU) & $f_{\rm merge}$\\
\hline
BH & $1$ & $2$   & uniform & $50$   & $0.287$ \\
BH & $1$ & $2$   & uniform & $200$  & $0.342$ \\
BH & $1$ & $2$   & uniform & $1000$ & $0.346$ \\
BH & $1$ & $2$   & thermal & $50$   & $0.358$ \\
BH & $2$ & $2$   & uniform & $50$   & $0.256$ \\
BH & $3$ & $2$   & uniform & $50$   & $0.233$ \\
BH & $4$ & $2$   & uniform & $50$   & $0.276$ \\
BH & $1$ & $3$   & uniform & $50$   & $0.304$ \\
NS & -   & $2$   & uniform & $50$   & $0.140$ \\
NS & -   & $2$   & uniform & $200$  & $0.158$ \\
NS & -   & $2$   & uniform & $1000$ & $0.152$ \\
NS & -   & $2$   & thermal & $50$   & $0.188$ \\
NS & -   & $1.5$ & uniform & $50$   & $0.141$ \\
\hline\hline
Binary Type & $\beta$ & $\alpha$ & $f(e)$ & $\amax$ (AU) & $f_{\rm merge}$\\
\hline
BH & $1$ & $2$   & uniform & -      & $0.045$ \\
BH & $2$ & $2$   & uniform & -      & $0.050$ \\
BH & $3$ & $2$   & uniform & -      & $0.036$ \\
BH & $4$ & $2$   & uniform & -      & $0.041$ \\
BH & $1$ & $3$   & uniform & -      & $0.051$ \\
NS & -   & $2$   & uniform & -      & $0.028$ \\
NS & -   & $1.5$ & uniform & -      & $0.032$ \\
\hline
\end{tabular}
\label{tab:models}
\end{table}

\subsection{Initial conditions}

In our study, we consider an SMBH orbited by a triple CO, made up of either three BHs or three NSs. We assume an SMBH mass of $4\times 10^6\msun$ for a Milky-Way like nucleus.

For the assumed distribution of BH masses, the initial mass function in triples is highly uncertain (and also likely to vary depending on the formation mechanism). Therefore we use a generic, parameterized form corresponding to a negative power-law distribution
\begin{equation}
\frac{dN}{dm} \propto M^{-\beta}
\label{eqn:bhmassfunc}
\end{equation}
in the mass range $5\msun$--$100\msun$\footnote{Note that pulsational pair instabilities may limit the maximum mass to $\sim 50\msun$ \citep{bel2016}.}, and sample the three BH masses independently. To study how the results depend on our assumptions, we vary the slope of the BH mass function in the range $\beta=1$, $2$, $3$, $4$ \citep{olea16}. For NSs, we fix the mass to $1.3\msun$ \citep*[e.g.][]{latt2005,fpb18}.

The numbers and spatial profiles of BHs and NSs surrounding SMBHs are unconstrained from an observational point of view. Stars and COs tend to form a power-law density cusp ($n(r)\propto r^{-\alpha}$) around an SMBH, where lighter and heavier objects develop shallower and steeper cusps, respectively \citep{bahcall76}. Typically, stars tend to have $\alpha\sim 1.5$--$1.75$, whereas BHs tend to have $\alpha\sim 2$--$3$ due to mass segregation \citep{alex17}. We assume that the BH and NS number densities follow a cusp with $\alpha=2$, and we study the effects of the cusp slope, by considering a steeper cusp ($\alpha=3$) for BHs, and a shallower cusp ($\alpha=1.5$) for NSs. We take the maximum outer semi-major axis to be $0.1$~pc\footnote{This value corresponds to the distance at which the octupole timescale is equal to the timescale on which accumulated fly-bys from stars tend to unbind the binary \citep{hoan18}.} \citep{hoan18}, and sample the outer orbital eccentricity from a thermal distribution \citep{jeans1919}.

The triple (BH-BH-BH or NS-NS-NS) inner and outer semi-major axis and eccentricity distributions are not well known observationally. Apart from their initial distributions, it should be taken into account that the dense environments characteristic of galactic nuclei or star clusters could cause both distributions to change or diffuse over time \citep{hop09}. We sample the triple inner and outer semi-major axis $\ainin$ and $\ainout$, respectively, by adopting a log-uniform distribution. Given the uncertainties in the orbital parameters of the triples, we explore a wide range of values for the outer triple semi-major axis; in particular, we study models with $a_{\rm 3,max}=50$ AU--$200$ AU--$1000$ AU. For the eccentricity, we consider models where we sample both $e_{12}$ and $e_3$ from uniform and thermal distributions. This is done to roughly reflect the different formation pathways of CO triples. Though the eccentricity mostly depends on the natal-kick and/or the common-envelope phase, it also depends on scattering and perturbations of the CO binaries by local COs and stars. As a consequence, we expect that \textit{in-situ} formation should naively favour circular systems, while CO triples that migrate from farther out should most likely prefer a thermal distribution, as a consequence of energy and angular momentum exchange through many dynamical encounters.

The relative inclinations of the inner and outer orbits are also poorly constrained observationally. In the field, observations have shown that inner and outer orbital planes of triples composed of massive stars tend to be inclined relative to each other \citep{moe18}, which may hint at moderately high relative inclinations between the inner and outer orbital planes of the triple COs. For the sake of generality, we draw the mutual inclination angles between the inner binary orbit with respect to the SMBH orbit ($i$) and the third companion in the triple's orbit ($i_3$) from an isotropic distribution. 

After we sample from the relevant distributions, we check that the \citet{mar01} criterion, namely
\begin{equation}
\frac{R_{\rm p,per}}{a_{\rm in}}\geq 2.8 \left[\left(1+\frac{m_{\rm per}}{m_{\rm in,tot}}\right)\frac{1+e_{\rm per}}{\sqrt{1-e_{\rm per}}} \right]^{2/5}\left(1.0-0.3\frac{I}{\pi}\right)\ ,
\label{eqn:stabts}
\end{equation}
is satisfied for each hierarchy in our four-body system in order to ensure orbital stability. In the previous equation, $m_{\rm in,tot}$ and $m_{\rm per}$ are the total inner mass and the perturber mass, respectively, $R_{\rm p,per}$ is the pericenter of the perturber, $a_{\rm in}$ is the inner semi-major axis, and $I$ (either $i$ or $i_3$ in our system) is the relative orbital inclination (in radians).

Given the above sets of initial parameters, we integrate the system of differential equations of motion of the four bodies
\begin{equation}
{\ddot{\textbf{r}}}_i=-G\sum\limits_{j\ne i}\frac{m_j(\textbf{r}_i-\textbf{r}_j)}{\left|\textbf{r}_i-\textbf{r}_j\right|^3}\ ,
\end{equation}
with $i=1$,$2$,$3$,$4$. The integrations are performed using the \textsc{archain} code \citep{mik06,mik08}, a fully regularized code able to model the evolution of systems of arbitrary mass ratios and large eccentricities with extreme accuracy. \textsc{archain} includes PN corrections up to order PN2.5. As a criterion for merger, we require that the relative distance between the COs is smaller than three times the sum of their relative radii.

For each of the $13$ different models (see Tab.~\ref{tab:models}), we run $\sim 800$--$1000$ realizations, for a total of $\sim 11000$ simulations. We fix the total integration time to $T=1$ Myr. Over this timescale, VRR may efficiently operate and reorient the binary centre-of-mass orbital plane with respect to the SMBH, thus affecting the relative inclination of the inner and outer orbital planes (we do not take into account VRR in our simulations). In this regard, VRR can affect the KL evolution (of the outer binary of the CO triple) even if the KL timescale is significantly shorter than $T_{\rm VRR}$. As shown in \citet{hamer18}, the distribution of maximum eccentricities is already significantly different (higher eccentricities) if $T_{\rm LK}/T_{\rm VRR}\sim 0.01$. This also affects the relative LK dynamics and renders the $4$-body approximation insufficient. The in-plane precession induced by the nuclear cluster potential and departures from spherical symmetry in the galactic nucleus would make the CO center of mass orbit precess even faster than is quantified by vector resonant relaxation alone in a MW-like nucleus \citep{petr17}. However, a relevant constraint for the triples lifetime in galactic nuclei comes from their evaporation as a result of the encounters with other stars, which takes place on timescales $\lesssim$ Myr for a Milky Way-like galaxy. Therefore, our final criterion for the total integration time is
\begin{equation}
T=\min(T_{\rm EV},1\ \mathrm{Myr})\ .
\end{equation}

In our simulations, the CO triple can undergo several fates: (i) the inner binary of the CO triple merges producing a GW merger event; (ii) the CO triple survives with orbital elements perturbed with respect to the initial ones; (iii) the CO triple is tidally broken apart by differential forces exerted by the SMBH, and its components will either be captured by the SMBH or ejected from the galactic nucleus. In particular, the latter can be caused by KL cycles of the outer orbit of the BH/NS triple, which becomes highly eccentric and destabilizes eventually disrupting the triple. This type of evolution was recently found to be relevant in population synthesis studies of 3+1 quadruples \citep{hamers2018,hamers2019}.

\begin{figure} 
\centering
\includegraphics[scale=0.55]{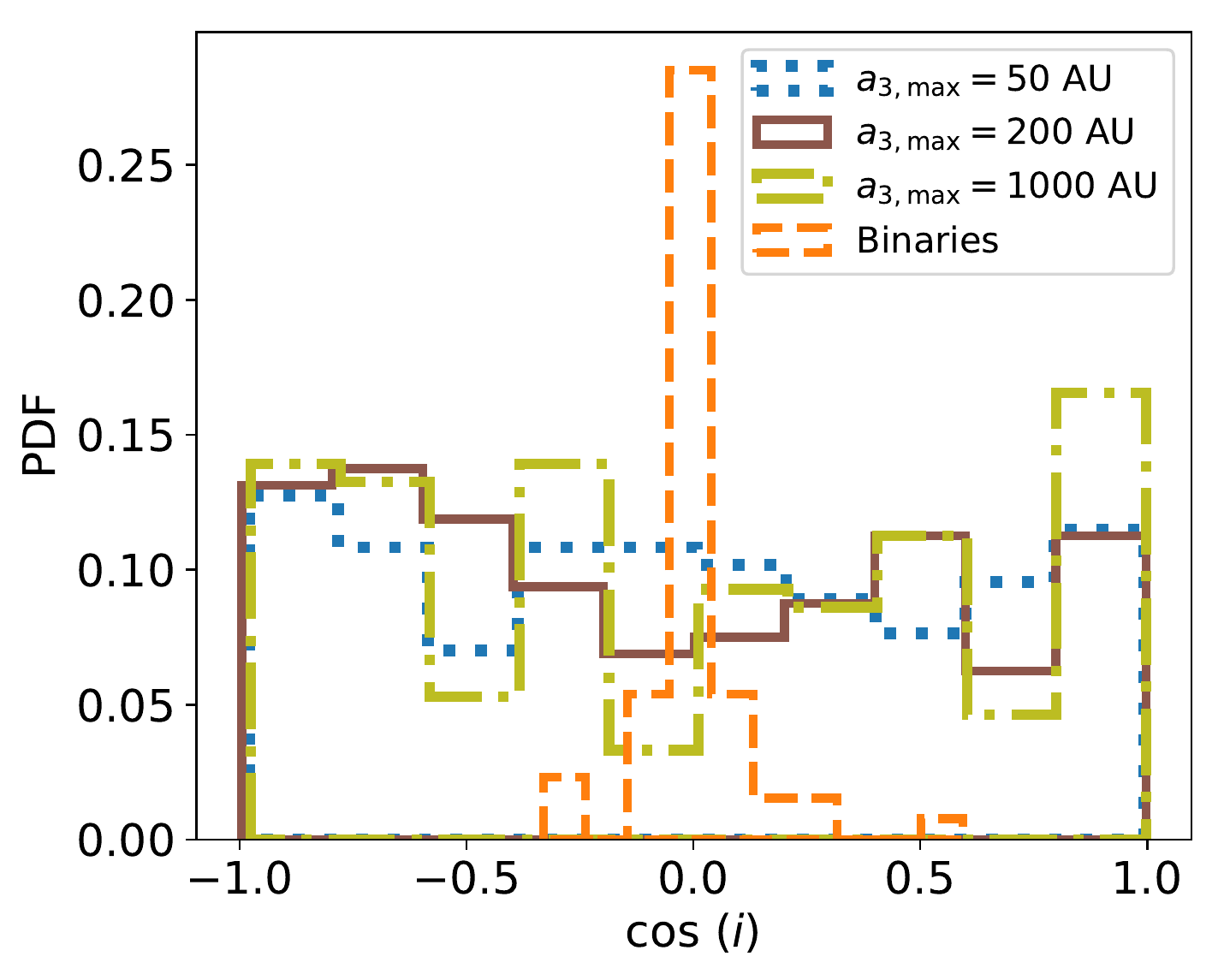}
\includegraphics[scale=0.55]{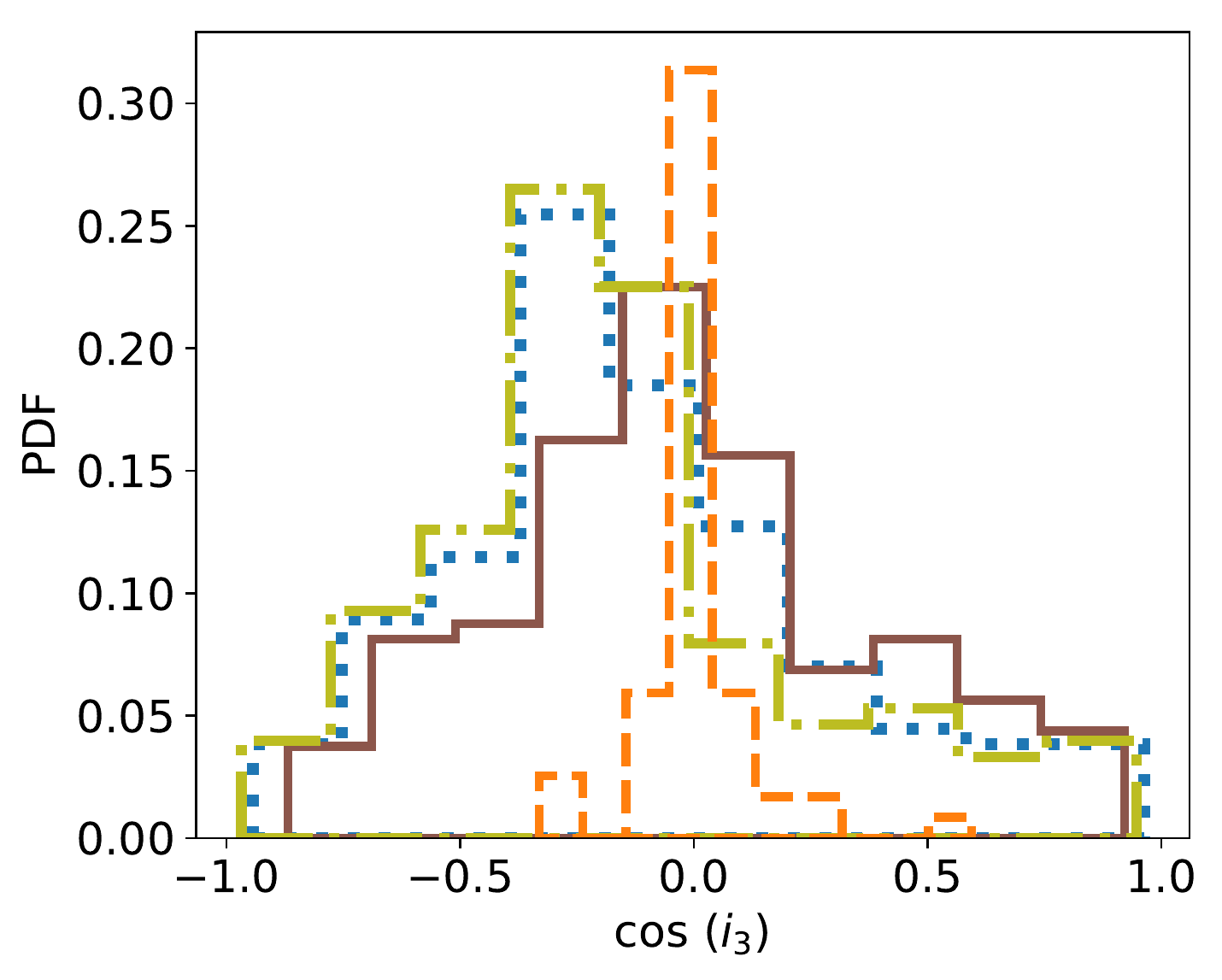}
\caption{Distributions of the initial orbital inclinations of the BH-BH binaries that merge in our simulations with respect to the SMBH ($i$, {\em top} panel) and the third companion in the triple ($i_3$, {\em bottom} panel) in Models MW, for $\beta=1$, $\alpha=2$ and different assumptions for $\amax$. Also shown is the distribution of initial inclinations for BH binary mergers \citep{fragrish2018}.}
\label{fig:angles}
\end{figure}

\begin{figure*} 
\centering
\includegraphics[scale=0.55]{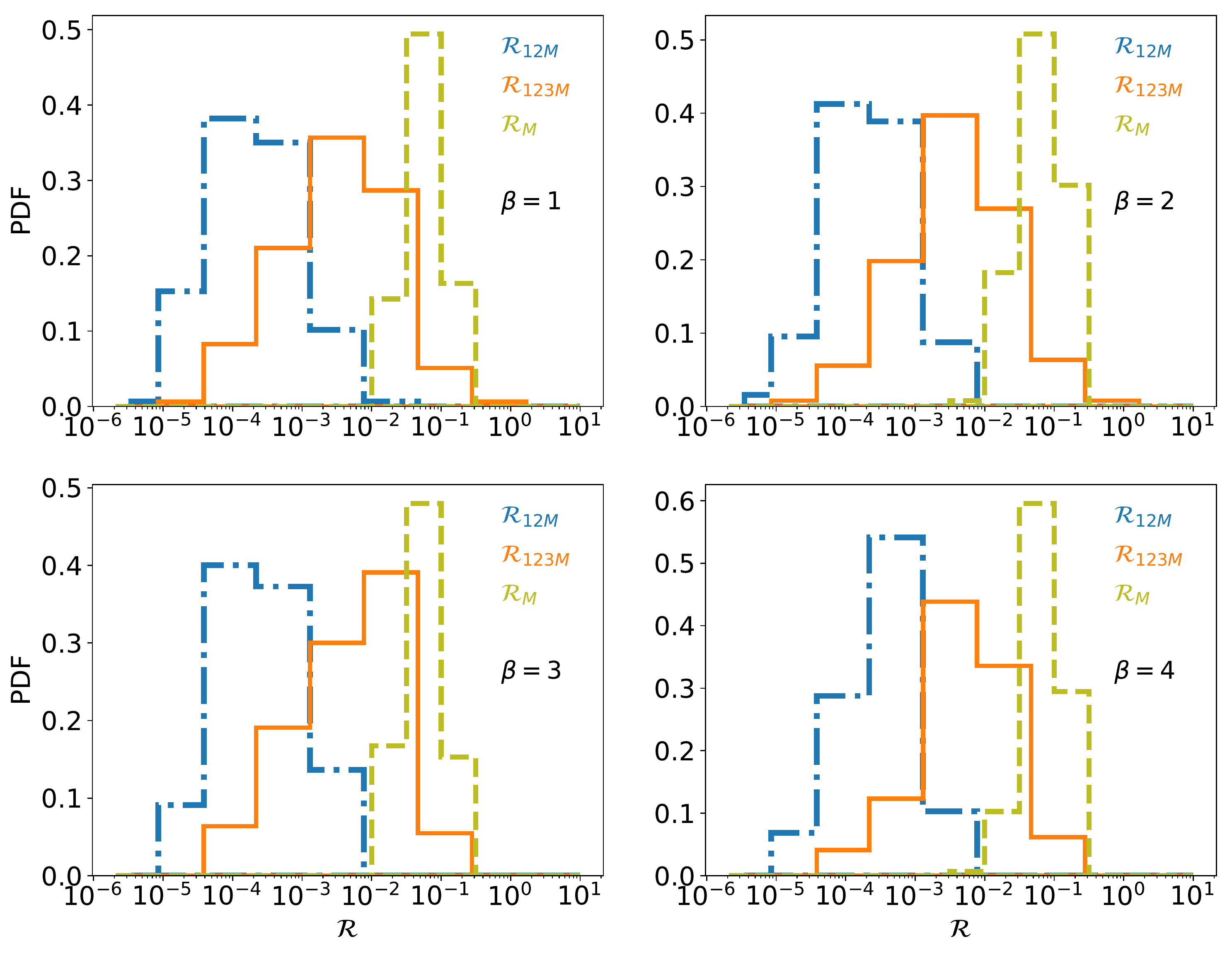}
\caption{Distributions of the ratios $\mathcal{R}_{12M}=T_{KL}^{123}/T_{KL}^{12M}$ and $\mathcal{R}_{123M}=T_{KL}^{123}/T_{KL}^{123M}$ of the BH-BH binaries that merge in our simulations in Models MW, $\alpha=2$, $\amax=50$ AU and different values of $\beta$.}
\label{fig:ratios123m}
\end{figure*}

\begin{figure} 
\centering
\includegraphics[scale=0.55]{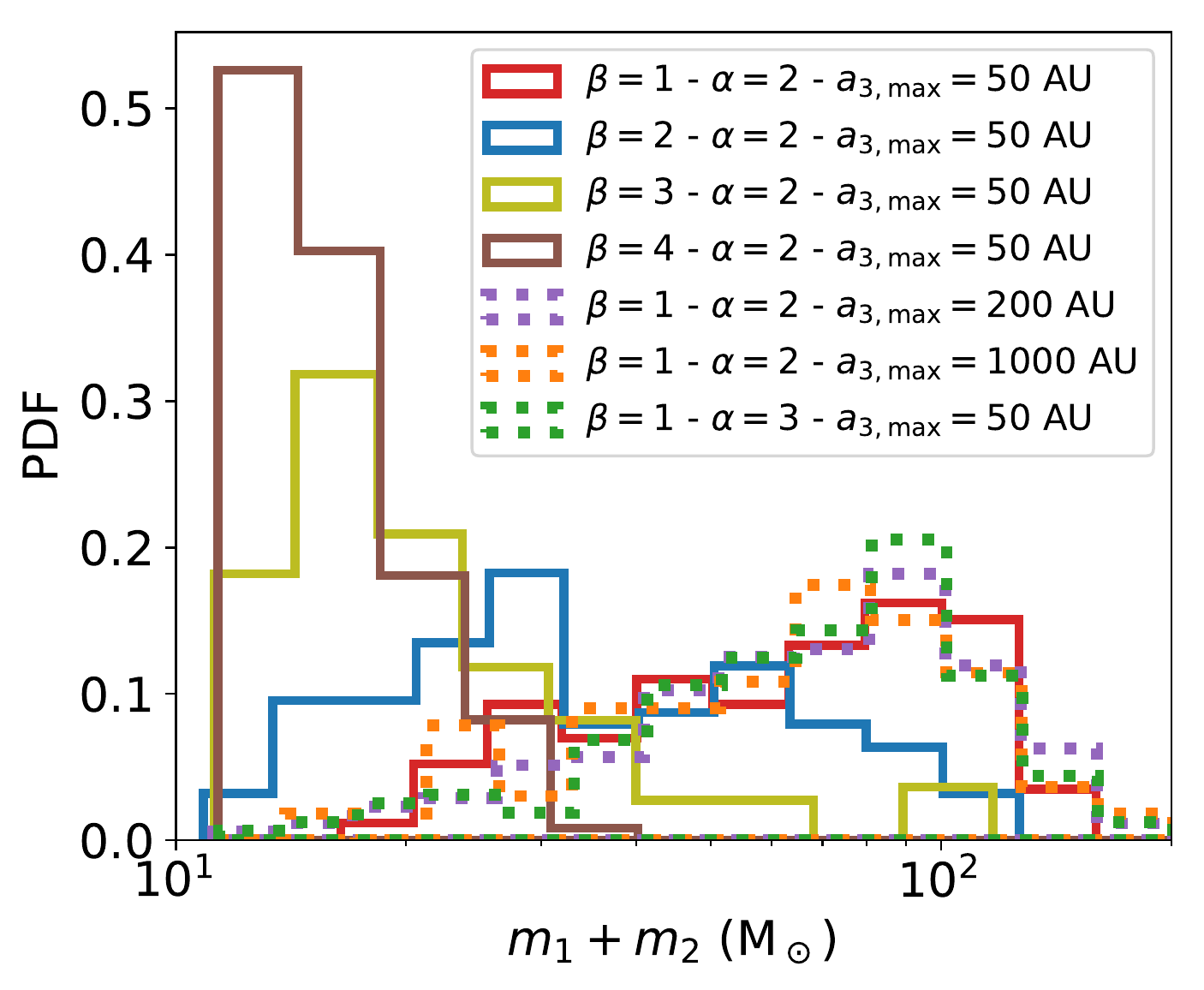}
\includegraphics[scale=0.55]{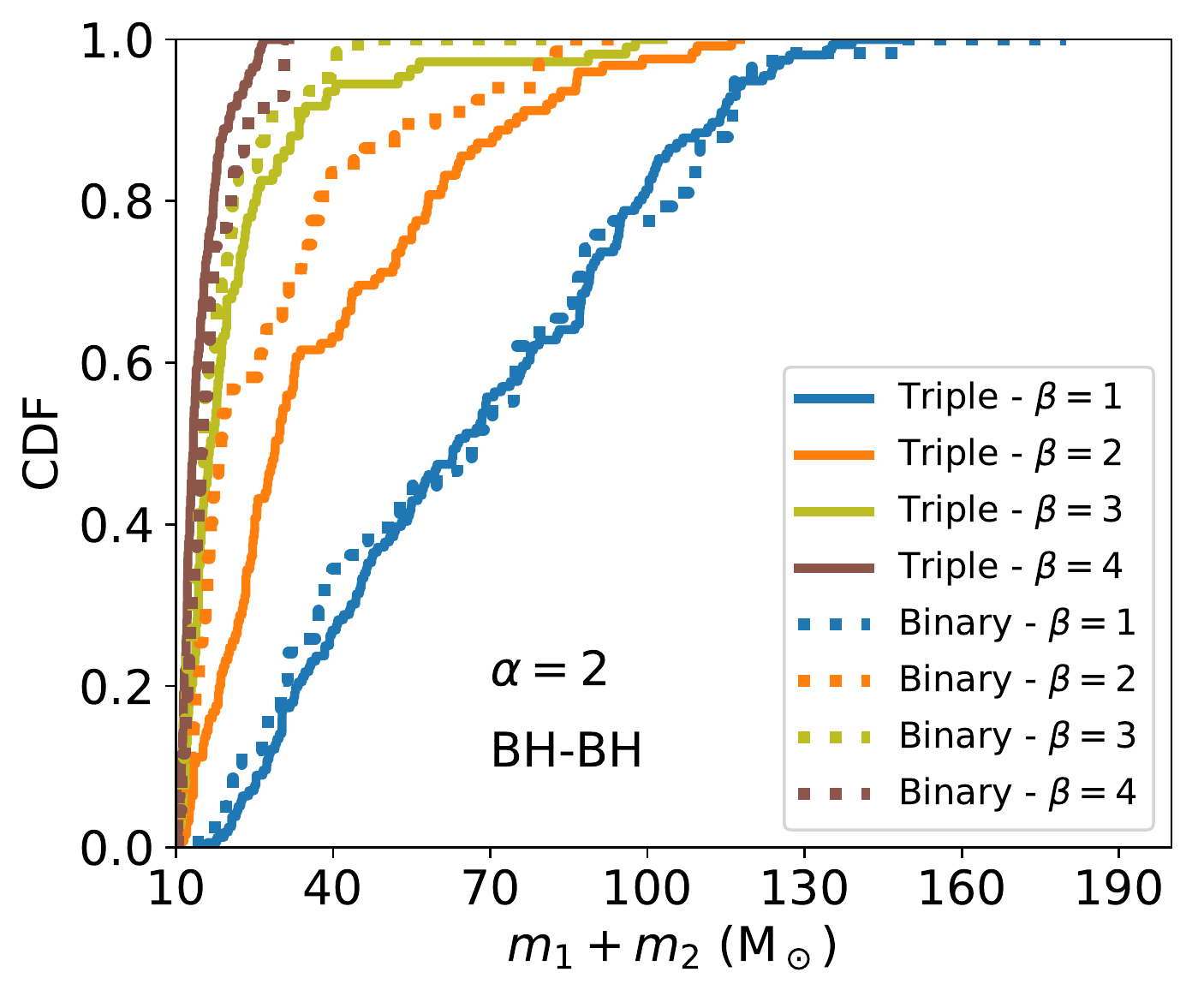}
\caption{\textit{Top}: mass distribution of merging BH-BH binaries in triples orbiting a $4 \times 10^6 \msun$ SMBH in a Milky Way-like nucleus, for different values of $\alpha$, $\beta$ and $\amax$. \textit{Bottom}: Comparison of the total mass of merging BHs in binaries \citep{fragrish2018} and triples, for different values of $\alpha=2$, $\beta$, and $\amax=50$ AU.}
\label{fig:massdist}
\end{figure}

\subsection{Inclination distribution}

We show in Fig.~\ref{fig:angles} the distribution of the initial orbital inclinations $i$ of the inner binaries that merge in our simulations with respect to the SMBH (top panel), for $\beta=1$, $\alpha=2$ and different $\amax$. For the case of a \textit{binary} CO orbiting an SMBH, most of the binaries that merge have initial inclinations $i\sim 90^\circ$ with respect to the SMBH, where the enhancement of the maximum eccentricity is expected to be the largest due to KL oscillations whenever not suppressed by GR precession \citep{fragrish2018}. The distribution thus results in a sharp peak at $\sim 90^\circ$ and only a few mergers have larger or smaller initial inclinations. For a \textit{triple} CO orbiting an SMBH, this distribution is no longer sharply peaked, but instead is almost isotropic.

\begin{figure} 
\centering
\includegraphics[scale=0.55]{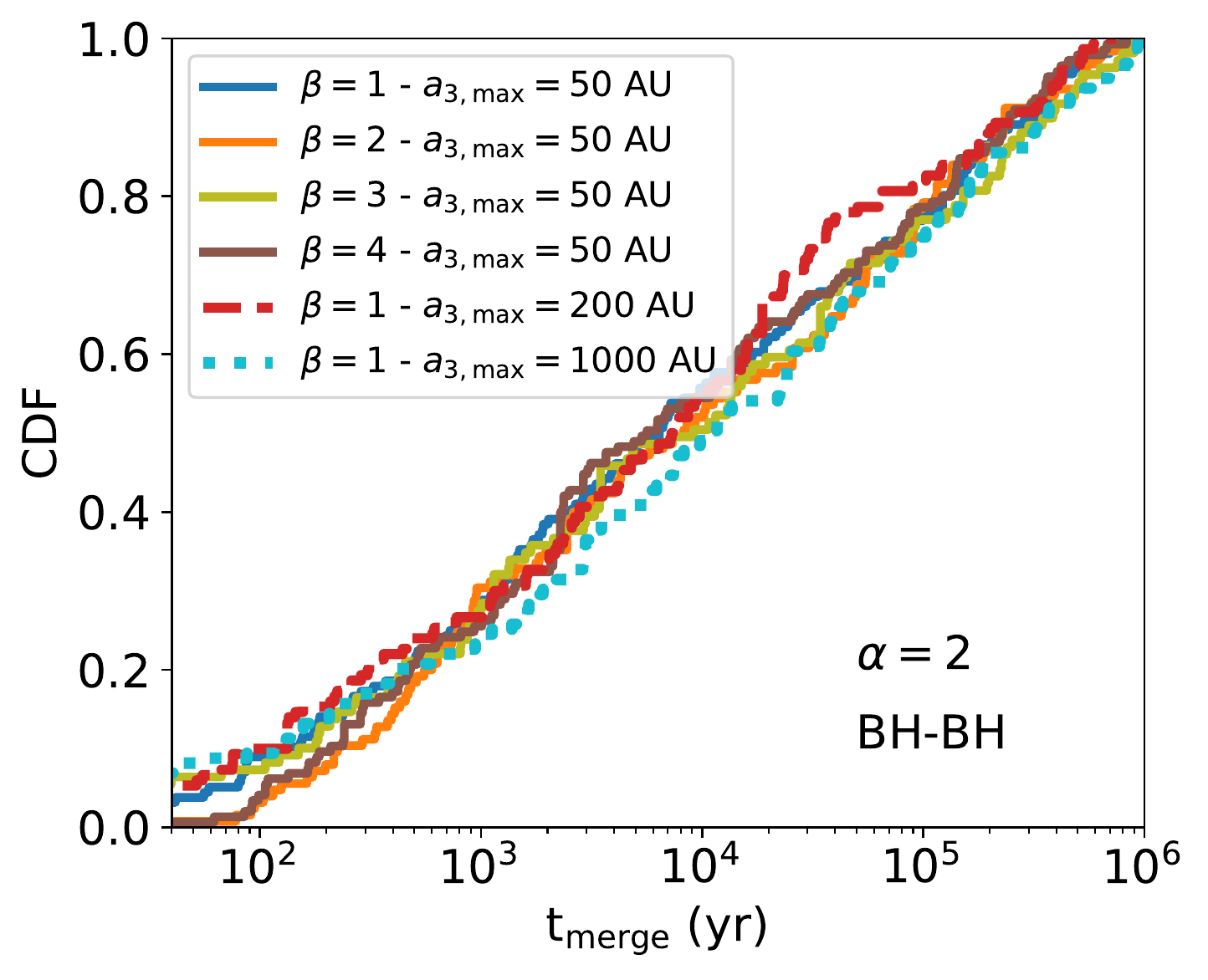}
\includegraphics[scale=0.55]{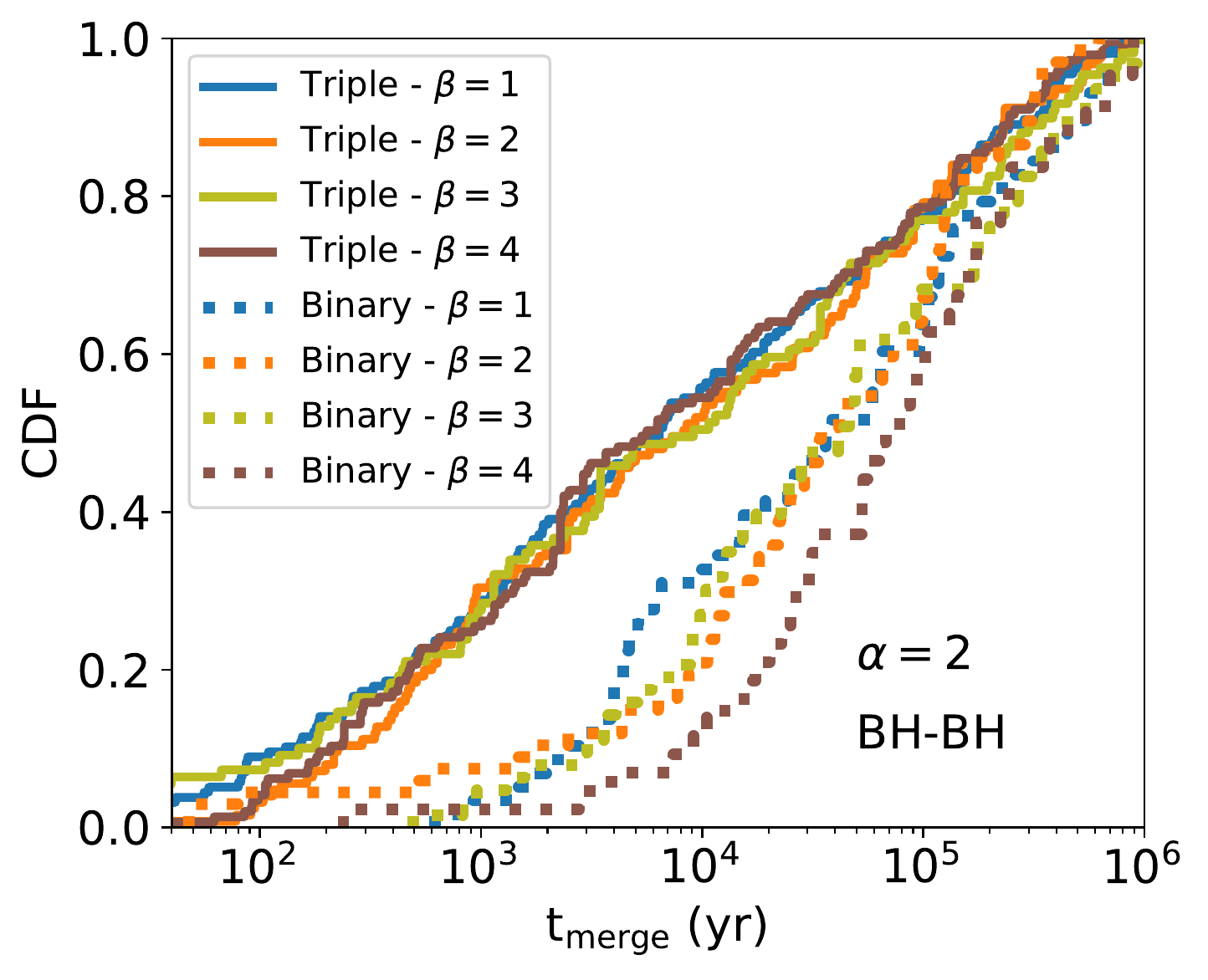}
\caption{Cumulative distribution functions of merger times for BH-BH binaries in triples orbiting an SMBH. Top: merger times of BH-BH mergers for different values of $\beta$ and $\amax$ (we fix $\alpha=2$). Bottom: comparison to mergers in binaries \citep{fragrish2018} for BH-BH mergers ($\amax=50$ AU, $\alpha=2$).}
\label{fig:ratesall}
\end{figure}

On the other hand, the distribution of the orbital inclination of the third companion with respect to the inner binary in the triple $i_3$ (bottom panel) is found to peak at $\sim 100^\circ$, but with non-negligible tails. For comparison, isolated triples that merge due to the KL mechanism show a very pronounced peak at $\sim 90^\circ$, with only a few mergers happening in low-inclination systems \citep{ant17}. There are two possible caveats to this. First, possible resonances between nodal precession and KL oscillations can arise. As shown in \citet{hamerslai2017}, this could make even low-inclination systems merge. Second, there are three different KL mechanisms competing, thus the eccentricity of the inner binary of the CO triple can be pumped up by the torque either of the SMBH or the third companion in the CO triple, for which the KL timescale is shorter \citep[e.g.][]{hamers2015}. We define the following ratios 
\begin{equation}
\mathcal{R}_{12M}=\frac{T_{KL}^{123}}{T_{KL}^{12M}}
\end{equation}
\begin{equation}
\mathcal{R}_{123M}=\frac{T_{KL}^{123}}{T_{KL}^{123M}}
\end{equation}
\begin{equation}
\mathcal{R}_{M}=\frac{T_{KL}^{123M}}{T_{KL}^{12M}}\ ,
\end{equation}
which we plot in Fig.~\ref{fig:ratios123m} for the BH-BH binaries that merge in our simulations in Models MW, $\alpha=2$, $\amax=50$ AU and different values of $\beta$. Clearly, the shape of the three distributions does not depend on the assumed value of $\beta$, and we find that $\mathcal{R}_{12M}$, $\mathcal{R}_{123M}$ and $\mathcal{R}_{M}$ peak at $\sim 10^{-4}$, $\sim 5\times 10^{-3}$ and $\sim 10^{-1}$, respectively. We note that, however, KL cycles in a given CO triple may be inactive if the relative inclinations are not in the KL window. As a consequence, the initial dynamical evolution of the CO triple can ultimately be dictated by a KL cycle that takes place on a long timescale. The latter can in turn excite the relative inclination of one of the other orbits, which could activate KL cycles that take place on a shorter timescale. Thus, the CO triple may experience rather different and rich dynamical paths, which tend to lead to chaotic behaviour \citep*{grishlai2018}.

\subsection{Mass distribution}

Figure~\ref{fig:massdist} (top panel) illustrates the distribution of the total mass of merging BH-BH-BH triples orbiting a $4 \times 10^6 \msun$ SMBH in a Milky Way-like nucleus, for different values of $\alpha$, $\beta$ and $\amax$. For \textit{binaries} orbiting an SMBH \citep{fragrish2018}, the resulting total mass distribution is not significantly affected by the slope of the density profile $\alpha$ around the SMBH, with a roughly constant shape in the range $\sim 25\msun$--$125 \msun$ and a tail extending up to $\sim 180\msun$. Also, the extent of the triple does not influence the total mass distribution. The only parameter that sets the distribution properties is the slope $\beta$ of the BH mass function. As expected, we find that the shallower the slope of the BH mass function, the larger the typical total mass of merging BH-BH binaries. For $\beta=1$, the distribution is peaked at a total mass of $\sim 30\msun$ with a tail that extends up to $\sim 125\msun$. For larger values of $\beta$, the tail gradually disappears and the distribution is peaked at $\sim 15\msun$. We find that $\sim 90\%$ of the mergers have total masses $\lesssim 20\msun$, $\lesssim 30\msun$, $\lesssim 70\msun$, and $\lesssim 100\msun$ for $\beta=1$, $2$, $3$, and $4$, respectively.

In the bottom panel of Fig.~\ref{fig:massdist}, we show a comparison between the total mass of merging BHs in binaries \citep{fragrish2018} and triples orbiting a $4 \times 10^6 \msun$ SMBH, assuming $\alpha=2$ and different values for the parameter $\beta$ (for triples we fix $\amax=50$ AU). We find that the total mass of merging binaries is nearly independent of the multiplicity of the system. Thus, the main parameter that governs the shape of the final distribution for the total mass of merged BHs is the slope of the BH mass function.

\subsection{Merger times}

Due to KL cycles, the eccentricity of the inner binary in the triple can approach almost unity when $i_0\sim 90^\circ$ and the inner binary merger time becomes shorter due to efficient energy dissipation at pericentre via KL oscillations. In the top panel of Fig.~\ref{fig:ratesall}, we show the cumulative distribution functions (CDFs) of merger times for BH-BH binaries in triples orbiting an SMBH of mass $\mmbh=4\times 10^6\msun$. Different values for the parameters $\alpha$, $\beta$ and $\amax$ do not affect the distribution of merger times.

In the bottom panel of Fig.~\ref{fig:ratesall}, we compare the CDFs of BHs merging in binaries \citep{fragrish2018} to those in triples orbiting a Milky Way-like SMBH, for different values of $\beta$. BHs merge in triples on shorter timescales compared to BHs that merge in binaries, independent of the slope of the BH mass function ($\amax=50$ AU, $\alpha=2$). We find that $\sim 50\%$ of the mergers have typical merger timescales of $\lesssim 3\times 10^3$ yr and $\lesssim 3\times 10^4$ yr, respectively, for the binary and triple cases. The most likely reason for these deviations comes from a richer dynamical evolution, with three KL cycles competing, combined with possible resonances that can arise in the case of triples orbiting an SMBH.

\subsection{Eccentricity in the LIGO band}

\begin{figure} 
\centering
\includegraphics[scale=0.55]{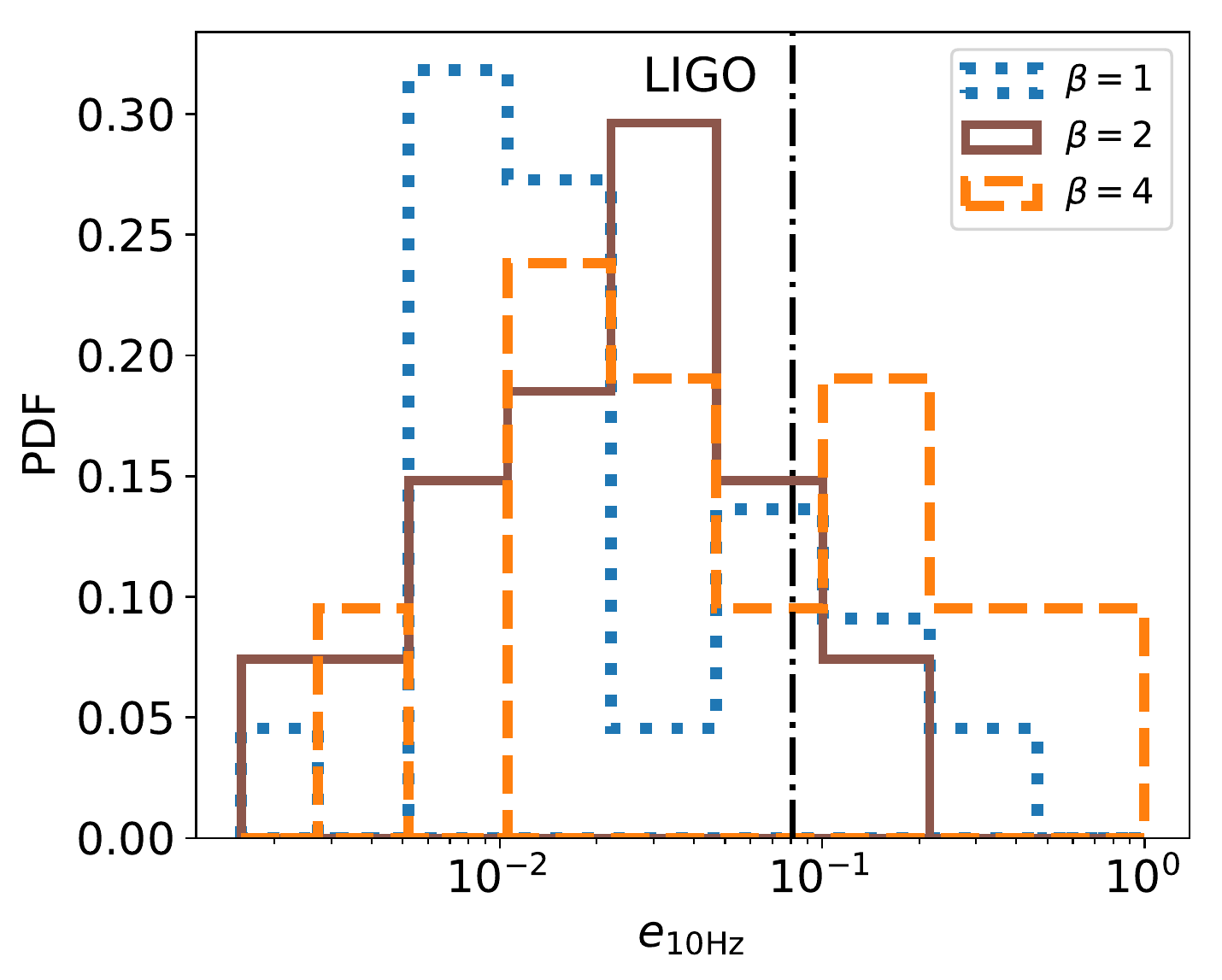}
\caption{Distribution of eccentricities at the moment the CO binaries enter the LIGO frequency band ($10$ Hz) for mergers produced by triples. The vertical line shows the minimum $e_{\rm 10Hz}=0.081$ where LIGO/VIRGO/KAGRA network may distinguish eccentric sources from circular sources \citep{gond2019}. A significant fraction of binaries formed in triples have a significant eccentricity in the LIGO band.}
\label{fig:ecc}
\end{figure}

Hierarchical configurations are expected to have eccentricities when entering the LIGO band ($10$ Hz) that are larger than for binaries that merge in isolation \citep[see e.g.][]{frl2019}. For the CO binaries that merge in our simulations, we compute a proxy for the GW frequency, i.e. the frequency corresponding to the harmonic that gives the maximum GW emission \citep{wen03}
\begin{equation} 
f_{\rm GW}=\frac{\sqrt{G(m_{11}+m_{12})}}{\pi}\frac{(1+e_{12})^{1.1954}}{[a_{12}(1-e_{12}^2)]^{1.5}}\ .
\end{equation}
In Figure~\ref{fig:ecc}, we illustrate the distribution of eccentricities at the moment the CO binaries enter the LIGO frequency band. We also plot the minimum $e_{\rm 10Hz}=0.081$ where the LIGO/VIRGO/KAGRA network may distinguish eccentric sources from circular sources \citep{gond2019}. A large fraction of systems that merge have a significant eccentricity in the LIGO band, compared to binaries that merge in isolation. We note that a similar signature could be found in CO binaries that merge near intermediate-mass black holes \citep{fragbr2019}, in the GW capture scenario in star clusters \citep{sam2018}, and in isolated hierarchical triples \citep{ant17} and quadruples \citep{fragk2019}.

\section{Discussion and conclusions}
\label{sect:conc}

We have presented a new scenario for merging BHs and NSs in galactic nuclei. We have studied the mergers of binary BHs and NSs in triples orbiting an SMBH in a Milky Way-like galactic nucleus. By conducting a series of high-precision numerical integrations, we have quantified the relative differences with the case where the BHs and NSs are in binaries \citep{fragrish2018}. We find that, unlike the binary case in which the initial orbital inclination of the merging BHs/NSs is sharply peaked at $\sim 90^\circ$, in the case of triples the distribution is significantly broader, due to the more complex dynamics of the triple-SMBH system, with three KL cycles competing, and possible resonances between nodal precession and KL oscillations. 

To accurately determine the global BH merger rate from this channel, we would need to quantify the population of triple BHs and NSs in galactic nuclei and star clusters, which is highly uncertain. For the binary case, this rate is thought to be of the order of $\sim 0.1$--$1\rm{\ Gpc}^{-3}\rm{\ yr}^{-1}$ \citep{antoper12,fragrish2018,hamer18,hoan18}. Nevertheless, the multiplicity fraction of high-mass main-sequence stars is large, with each star having two or more companions on average \citep{sana2017}. We find that the merger fraction in triple BH and NS systems can be larger than the one for binaries, by up to a factor of $\sim 8$ (see the last column of Tab.~\ref{tab:models}). Our results thus suggest that dynamically-driven BH and NS mergers in this scenario may be important and contribute to the merger events observed by LIGO/VIRGO. Moreover, as for other hierarchical systems, we have found that binaries merging in CO triples are expected to have eccentricities in the LIGO band ($10$ Hz) larger than binaries that merge in isolation, with a peak at $\sim 1$.

We note that in the case the inner binary merges, we stop our simulations. In reality, however, the merger remnant should be imparted a recoil velocity kick that could unbind the system \citep[e.g.][]{lou2010}, now a binary comprised of the merger remnant itself and the third companion in the former triple. If the kick does not unbind the system, a second-generation binary is born that can undergo KL oscillations as a result of the SMBH tidal field, which can eventually result in a second merger. Therefore, the presence of the massive SMBH could further enhance the merger rate from this scenario, by a factor of up to $\sim 2$.

We caution, however, that there are a lot of uncertainties regarding the exact population of stars and CO triples, as well as their mass and orbital parameters distributions. Moreover, we have not taken into account the pre-CO evolution, which could be relevant in determining the properties of triple COs \citep{hamers2013,shapp2013,ant17,toonen2018}. In our simulations, we only consider triples made up of all BHs or all NSs, which implies that their progenitors had a mass ratio near unity. While this is the case for a subset of systems, additional simulations incorporating all possible compositions of the CO triples will be needed to precisely compute how the merger fraction depends on the specific composition of the triple. Importantly, the maximum outer semi-major axis of triples plays an important role in determining the relevant timescales. In particular, both the the 3+1 interaction timescale and the triple evaporation timescale depend crucially on it. As we show, these timescales can be as short as a few Myrs, such that that more compact triples will probably dominate the rate. This would in turn imply that CO triples in galactic nuclei need to be born more compact than in the field, if they were to contribute to the observed GW merger events.

What makes these events	especially interesting for the GW community is the fact that they have enhanced chance of being detected also in the electromagnetic (EM) spectrum. For a binary NS merger, the intensity of the gamma-ray radiation, which primarily depends on the amount of mass available for accretion, is primarily determined by the properties (namely mass and equation of state) of the merging stars \citep{Shibata2006,Rezzolla2010,Giacomazzo2013,Hotokezaka2013}. However, if	the merger event drives a relativistic jet as evinced by the observations of GW170817/GRB170817A \citep{Abbott2017a} and detailed broadband modeling \citep{Lazzati2018}, then the density $n$ of the medium in which the event occurs also plays an important role. In fact, the relativistic jet will produce bright radiation (known as an afterglow), spanning the full EM spectrum, from hard X-rays to the radio, and the peak of this emission scales as $\propto n^{1/2}$ \citep{Sari1998}. The binary NS mergers from triples orbiting an SMBH studied here typically have short merger times, even shorter than those of isolated binaries alone, as studied by \citet{fragrish2018}. As such, they are expected to merge in the innermost regions of a galaxy, where densities	are at their highest. For	gas densities of a few cm$^{-3}$, the EM brightness would allow for their detection in multiple bands well beyond the LIGO horizon, as computed by \citet{fragrish2018}. EM observations are a crucial element allowing for event localization, and hence to test the nuclear origins of these events.

Mergers	of binaries made up of two BHs are more	likely to be dark, lacking the natural reservoir of accretion matter provided by the tidal disruption of a NS immediately preceding the final coalescence. While some	reservoir of material could still be found around one of the BHs as a dead disk remnant from the SN explosion \citep{Perna2016}, this may not be a common scenario. If the SMBH is surrounded by a dense disk as in active galactic nuclei then, if the binary BH merges within the disk, a temporary accretion of matter at high rates can ensue \citep{Bartos2017}, resulting in a possible EM counterpart. Nevertheless, for the case of a binary BH in a triple there is the additional possibility of producing electromagnetic radiation if the third object, instead of being a BH, is a main sequence star that fills its Roche lobe and generates a circumbinary disk around the inner binary BH \citep{Chang2018}. As the binary BH merges, the loss of energy via GWs along with a possible merger kick can result in orbital crossings of the disk particles, resulting in disk shocking. Alternatively (or additionally) the circumbinary disk will eventually accrete on a viscous timescale upon merger of the binary. All together, binary mergers from triples near SMBHs may constitute an especially important channel for combined GW/EM transients.

\section*{Acknowledgements}

We thank the anonymous referee for useful comments. GF is supported by the Foreign Postdoctoral Fellowship Program of the Israel Academy of Sciences and Humanities. GF also acknowledges support from an Arskin postdoctoral fellowship. NL and RP acknowledge support by NSF award AST-1616157. GF thanks Seppo Mikkola for helpful discussions on the use of the code \textsc{archain}. Simulations were run on the \textit{Astric} cluster at the Hebrew University of Jerusalem.
The Center for Computational Astro-physics at the Flatiron Institute is supported by the Simons Foundation.

\bibliographystyle{mn2e}
\bibliography{refs}

\begin{thebibliography}{}

\bibitem[\protect\citeauthoryear{{Abbott}, {Abbott}, {Abbott}, {Acernese},
  {Ackley}, {Adams}, {Adams}, {Addesso}, {Adhikari}, {Adya} \& et al.}{{Abbott}
  et~al.}{2017}]{Abbott2017a}
{Abbott} B.~P.,  {Abbott} R.,  {Abbott} T.~D.,  {Acernese} F.,  {Ackley} K.,
  {Adams} C.,  {Adams} T.,  {Addesso} P.,  {Adhikari} R.~X.,  {Adya} V.~B.,
  et al. 2017, \apjl, 848, L13

\bibitem[\protect\citeauthoryear{{Alexander}}{{Alexander}}{2017}]{alex17}
{Alexander} T.,  2017, Ann Rev Astron Astrop, 55, 17

\bibitem[\protect\citeauthoryear{{Antognini}, {Shappee}, {Thompson} \&
  {Amaro-Seoane}}{{Antognini} et~al.}{2014}]{antognini14}
{Antognini} J.~M.,  {Shappee} B.~J.,  {Thompson} T.~A.,    {Amaro-Seoane} P.,
  2014, \mnras, 439, 1079

\bibitem[\protect\citeauthoryear{{Antognini}}{{Antognini}}{2015}]{antognini15}
{Antognini} J.~M.~O.,  2015, \mnras, 452, 3610

\bibitem[\protect\citeauthoryear{{Antonini}, {Chatterjee}, {Rodriguez},
  {Morscher}, {Pattabiraman}, {Kalogera} \& {Rasio}}{{Antonini}
  et~al.}{2016}]{antcha2016}
{Antonini} F.,  {Chatterjee} S.,  {Rodriguez} C.~L.,  {Morscher} M.,
  {Pattabiraman} B.,  {Kalogera} V.,    {Rasio} F.~A.,  2016, \apj, 816, 65

\bibitem[\protect\citeauthoryear{{Antonini} \& {Perets}}{{Antonini} \&
  {Perets}}{2012}]{antoper12}
{Antonini} F.,  {Perets} H.~B.,  2012, \apj, 757, 27

\bibitem[\protect\citeauthoryear{{Antonini} \& {Rasio}}{{Antonini} \&
  {Rasio}}{2016}]{ant16}
{Antonini} F.,  {Rasio} F.~A.,  2016, \apj, 831, 187

\bibitem[\protect\citeauthoryear{{Antonini}, {Toonen} \& {Hamers}}{{Antonini}
  et~al.}{2017}]{ant17}
{Antonini} F.,  {Toonen} S.,    {Hamers} A.~S.,  2017, \apj, 841, 77

\bibitem[\protect\citeauthoryear{{Askar}, {Szkudlarek}, {Gondek-Rosi\'{n}ska},
  {Giersz} \& {Bulik}}{{Askar} et~al.}{2017}]{askar17}
{Askar} A.,  {Szkudlarek} M.,  {Gondek-Rosi\'{n}ska} D.,  {Giersz} M.,
  {Bulik} T.,  2017, \mnras, 464, L36

\bibitem[\protect\citeauthoryear{{Bahcall} \& {Wolf}}{{Bahcall} \&
  {Wolf}}{1976}]{bahcall76}
{Bahcall} J.~N.,  {Wolf} R.~A.,  1976, \apj, 209, 214

\bibitem[\protect\citeauthoryear{{Banerjee}}{{Banerjee}}{2018}]{baner18}
{Banerjee} S.,  2018, \mnras, 473, 909

\bibitem[\protect\citeauthoryear{{Bartko} et~al.,}{{Bartko}
  et~al.}{2009}]{bart09}
{Bartko} H.,  et~al., 2009, \apj, 697, 1741

\bibitem[\protect\citeauthoryear{{Bartos}, {Kocsis}, {Haiman} \&
  {M{\'a}rka}}{{Bartos} et~al.}{2017}]{Bartos2017}
{Bartos} I.,  {Kocsis} B.,  {Haiman} Z.,    {M{\'a}rka} S.,  2017, \apj, 835,
  165

\bibitem[\protect\citeauthoryear{{Belczynski}, {Heger}, {Gladysz}, {Ruiter},
  {Woosley}, {Wiktorowicz}, {Chen}, {Bulik}, {O'Shaughnessy}, {Holz}, {Fryer}
  \& {Berti}}{{Belczynski} et~al.}{2016}]{bel2016}
{Belczynski} K.,  {Heger} A.,  {Gladysz} W.,  {Ruiter} A.~J.,  {Woosley} S.,
  {Wiktorowicz} G.,  {Chen} H.~Y.,  {Bulik} T.,  {O'Shaughnessy} R.,  {Holz}
  D.~E.,  {Fryer} C.~L.,    {Berti} E.,  2016, \aap, 594, A97

\bibitem[\protect\citeauthoryear{{Belczynski}, {Holz}, {Bulik} \&
  {O'Shaughnessy}}{{Belczynski} et~al.}{2016}]{bel16b}
{Belczynski} K.,  {Holz} D.~E.,  {Bulik} T.,    {O'Shaughnessy} R.,  2016,
  \nat, 534, 512

\bibitem[\protect\citeauthoryear{{Binney} \& {Tremaine}}{{Binney} \&
  {Tremaine}}{1987}]{binntrem87}
{Binney} J.,  {Tremaine} S.,  1987, {Galactic dynamics}

\bibitem[\protect\citeauthoryear{{Blaes}, {Lee} \& {Socrates}}{{Blaes}
  et~al.}{2002}]{blaes2002}
{Blaes} O.,  {Lee} M.~H.,    {Socrates} A.,  2002, \apj, 578, 775

\bibitem[\protect\citeauthoryear{{B{\"o}ker}, {Sarzi}, {McLaughlin}, {van der
  Marel}, {Rix}, {Ho} \& {Shields}}{{B{\"o}ker} et~al.}{2004}]{boker2004}
{B{\"o}ker} T.,  {Sarzi} M.,  {McLaughlin} D.~E.,  {van der Marel} R.~P.,
  {Rix} H.-W.,  {Ho} L.~C.,    {Shields} J.~C.,  2004, \aj, 127, 105

\bibitem[\protect\citeauthoryear{{Bouvier}, {Duch{\^e}ne}, {Mermilliod} \&
  {Simon}}{{Bouvier} et~al.}{2001}]{bouvier01}
{Bouvier} J.,  {Duch{\^e}ne} G.,  {Mermilliod} J.-C.,    {Simon} T.,  2001,
  \aap, 375, 989

\bibitem[\protect\citeauthoryear{{Bouvier}, {Rigaut} \& {Nadeau}}{{Bouvier}
  et~al.}{1997}]{bouvier97}
{Bouvier} J.,  {Rigaut} F.,    {Nadeau} D.,  1997, \aap, 323, 139

\bibitem[\protect\citeauthoryear{{Chang} \& {Murray}}{{Chang} \&
  {Murray}}{2018}]{Chang2018}
{Chang} P.,  {Murray} N.,  2018, \mnras, 474, L12

\bibitem[\protect\citeauthoryear{{Chu}, {Do}, {Hees}, {Ghez}, {Naoz}, {Witzel},
  {Sakai}, {Chappell}, {Gautam}, {Lu} \& {Matthews}}{{Chu}
  et~al.}{2018}]{chu2018}
{Chu} D.~S.,  {Do} T.,  {Hees} A.,  {Ghez} A.,  {Naoz} S.,  {Witzel} G.,
  {Sakai} S.,  {Chappell} S.,  {Gautam} A.~K.,  {Lu} J.~R.,    {Matthews} K.,
  2018, \apj, 854, 12

\bibitem[\protect\citeauthoryear{{Cohn}, {Lugger}, {Couch}, {Anderson}, {Cool},
  {van den Berg}, {Bogdanov}, {Heinke} \& {Grindlay}}{{Cohn}
  et~al.}{2010}]{cohn10}
{Cohn} H.~N.,  {Lugger} P.~M.,  {Couch} S.~M.,  {Anderson} J.,  {Cool} A.~M.,
  {van den Berg} M.,  {Bogdanov} S.,  {Heinke} C.~O.,    {Grindlay} J.~E.,
  2010, \apj, 722, 20

\bibitem[\protect\citeauthoryear{{C{\^o}t{\'e}}, {Piatek}, {Ferrarese},
  {Jord{\'a}n}, {Merritt}, {Peng}, {Ha{\c s}egan}, {Blakeslee}, {Mei}, {West},
  {Milosavljevi{\'c}} \& {Tonry}}{{C{\^o}t{\'e}} et~al.}{2006}]{cote2006}
{C{\^o}t{\'e}} P.,  {Piatek} S.,  {Ferrarese} L.,  {Jord{\'a}n} A.,  {Merritt}
  D.,  {Peng} E.~W.,  {Ha{\c s}egan} M.,  {Blakeslee} J.~P.,  {Mei} S.,  {West}
  M.~J.,  {Milosavljevi{\'c}} M.,    {Tonry} J.~L.,  2006, \apjs, 165, 57

\bibitem[\protect\citeauthoryear{{Dunstall}, {Dufton}, {Sana}, {Evans},
  {Howarth}, {Sim{\'o}n-D{\'{\i}}az}, {de Mink}, {Langer}, {Ma{\'{\i}}z
  Apell{\'a}niz} \& {Taylor}}{{Dunstall} et~al.}{2015}]{duns2015}
{Dunstall} P.~R.,  {Dufton} P.~L.,  {Sana} H.,  {Evans} C.~J.,  {Howarth}
  I.~D.,  {Sim{\'o}n-D{\'{\i}}az} S.,  {de Mink} S.~E.,  {Langer} N.,
  {Ma{\'{\i}}z Apell{\'a}niz} J.,    {Taylor} W.~D.,  2015, \aap, 580, A93

\bibitem[\protect\citeauthoryear{{Duquennoy} \& {Mayor}}{{Duquennoy} \&
  {Mayor}}{1991}]{duq91}
{Duquennoy} A.,  {Mayor} M.,  1991, A\& A, 248, 485

\bibitem[\protect\citeauthoryear{{Fan} et~al.,}{{Fan}  et~al.}{1996}]{fan96}
{Fan} X.,  et~al., 1996, AJ, 112, 628

\bibitem[\protect\citeauthoryear{{Fragione}}{{Fragione}}{2019}]{fra2019}
{Fragione} G.,  2019, arXiv e-prints, p. arXiv:1903.03117

\bibitem[\protect\citeauthoryear{{Fragione} \& {Bromberg}}{{Fragione} \&
  {Bromberg}}{2019}]{fragbr2019}
{Fragione} G.,  {Bromberg} O.,  2019, arXiv e-prints, p. arXiv:1903.09659

\bibitem[\protect\citeauthoryear{{Fragione}, {Grishin}, {Leigh}, {Perets} \&
  {Perna}}{{Fragione} et~al.}{2018}]{fragrish2018}
{Fragione} G.,  {Grishin} E.,  {Leigh} N.~W.~C.,  {Perets} H.~B.,    {Perna}
  R.,  2018, arXiv e-prints

\bibitem[\protect\citeauthoryear{{Fragione} \& {Kocsis}}{{Fragione} \&
  {Kocsis}}{2018}]{frak18}
{Fragione} G.,  {Kocsis} B.,  2018, Phys Rev Lett, 121, 161103

\bibitem[\protect\citeauthoryear{{Fragione} \& {Kocsis}}{{Fragione} \&
  {Kocsis}}{2019}]{fragk2019}
{Fragione} G.,  {Kocsis} B.,  2019, \mnras, 486, 4781

\bibitem[\protect\citeauthoryear{{Fragione} \& {Loeb}}{{Fragione} \&
  {Loeb}}{2019}]{frl2019}
{Fragione} G.,  {Loeb} A.,  2019, \mnras, 486, 4443

\bibitem[\protect\citeauthoryear{{Fragione}, {Pavl\'{i}k} \&
  {Banerjee}}{{Fragione} et~al.}{2018}]{fpb18}
{Fragione} G.,  {Pavl\'{i}k} V.,    {Banerjee} S.,  2018, \mnras, 480, 4955

\bibitem[\protect\citeauthoryear{{Fregeau}, {Ivanova} \& {Rasio}}{{Fregeau}
  et~al.}{2009}]{fregeau2009}
{Fregeau} J.~M.,  {Ivanova} N.,    {Rasio} F.~A.,  2009, \apj, 707, 1533

\bibitem[\protect\citeauthoryear{{Geller}, {Hurley} \& {Mathieu}}{{Geller}
  et~al.}{2013}]{geller13}
{Geller} A.~M.,  {Hurley} J.~R.,    {Mathieu} R.~D.,  2013, AJ, 145, 8

\bibitem[\protect\citeauthoryear{{Geller}, {Mathieu}, {Harris} \&
  {McClure}}{{Geller} et~al.}{2009}]{geller09}
{Geller} A.~M.,  {Mathieu} R.~D.,  {Harris} H.~C.,    {McClure} R.~D.,  2009,
  AJ, 137, 3743

\bibitem[\protect\citeauthoryear{{Giacomazzo}, {Perna}, {Rezzolla}, {Troja} \&
  {Lazzati}}{{Giacomazzo} et~al.}{2013}]{Giacomazzo2013}
{Giacomazzo} B.,  {Perna} R.,  {Rezzolla} L.,  {Troja} E.,    {Lazzati} D.,
  2013, \apjl, 762, L18

\bibitem[\protect\citeauthoryear{{Gond{\'a}n} \& {Kocsis}}{{Gond{\'a}n} \&
  {Kocsis}}{2019}]{gond2019}
{Gond{\'a}n} L.,  {Kocsis} B.,  2019, \apj, 871, 178

\bibitem[\protect\citeauthoryear{{Gond{\'a}n}, {Kocsis}, {Raffai} \&
  {Frei}}{{Gond{\'a}n} et~al.}{2018}]{gondan2018}
{Gond{\'a}n} L.,  {Kocsis} B.,  {Raffai} P.,    {Frei} Z.,  2018, \apj, 860, 5

\bibitem[\protect\citeauthoryear{{Grindlay}, {Bailyn}, {Cohn}, {Lugger},
  {Thorstensen} \& {Wegner}}{{Grindlay} et~al.}{1988}]{grindlay88}
{Grindlay} J.~E.,  {Bailyn} C.~D.,  {Cohn} H.,  {Lugger} P.~M.,  {Thorstensen}
  J.~R.,    {Wegner} G.,  1988, \apjl, 334, L25

\bibitem[\protect\citeauthoryear{{Grishin}, {Lai} \& {Perets}}{{Grishin}
  et~al.}{2018}]{grishlai2018}
{Grishin} E.,  {Lai} D.,    {Perets} H.~B.,  2018, \mnras, 474, 3547

\bibitem[\protect\citeauthoryear{{Grishin}, {Perets} \& {Fragione}}{{Grishin}
  et~al.}{2018}]{grish18}
{Grishin} E.,  {Perets} H.~B.,    {Fragione} G.,  2018, \mnras, 481, 4907

\bibitem[\protect\citeauthoryear{{Hailey}, {Mori}, {Bauer}, {Berkowitz}, {Hong}
  \& {Hord}}{{Hailey} et~al.}{2018}]{hailey18}
{Hailey} C.~J.,  {Mori} K.,  {Bauer} F.~E.,  {Berkowitz} M.~E.,  {Hong} J.,
  {Hord} B.~J.,  2018, \nat, 556, 70

\bibitem[\protect\citeauthoryear{{Hamers}}{{Hamers}}{2018}]{hamers2018}
{Hamers} A.~S.,  2018, \mnras, 478, 620

\bibitem[\protect\citeauthoryear{{Hamers}}{{Hamers}}{2019}]{hamers2019}
{Hamers} A.~S.,  2019, \mnras, 482, 2262

\bibitem[\protect\citeauthoryear{{Hamers}, {Bar-Or}, {Petrovich} \&
  {Antonini}}{{Hamers} et~al.}{2018}]{hamer18}
{Hamers} A.~S.,  {Bar-Or} B.,  {Petrovich} C.,    {Antonini} F.,  2018, \apj,
  865, 2

\bibitem[\protect\citeauthoryear{{Hamers} \& {Lai}}{{Hamers} \&
  {Lai}}{2017}]{hamerslai2017}
{Hamers} A.~S.,  {Lai} D.,  2017, \mnras, 470, 1657

\bibitem[\protect\citeauthoryear{{Hamers}, {Perets}, {Antonini} \& {Portegies
  Zwart}}{{Hamers} et~al.}{2015}]{hamers2015}
{Hamers} A.~S.,  {Perets} H.~B.,  {Antonini} F.,    {Portegies Zwart} S.~F.,
  2015, \mnras, 449, 4221

\bibitem[\protect\citeauthoryear{{Hamers}, {Pols}, {Claeys} \&
  {Nelemans}}{{Hamers} et~al.}{2013}]{hamers2013}
{Hamers} A.~S.,  {Pols} O.~R.,  {Claeys} J.~S.~W.,    {Nelemans} G.,  2013,
  \mnras, 430, 2262

\bibitem[\protect\citeauthoryear{{Hoang}, {Naoz}, {Kocsis}, {Rasio} \&
  {Dosopoulou}}{{Hoang} et~al.}{2018}]{hoan18}
{Hoang} B.-M.,  {Naoz} S.,  {Kocsis} B.,  {Rasio} F.~A.,    {Dosopoulou} F.,
  2018, \apj, 856, 140

\bibitem[\protect\citeauthoryear{{Hopman}}{{Hopman}}{2009}]{hop09}
{Hopman} C.,  2009, \apj, 700, 1933

\bibitem[\protect\citeauthoryear{{Hotokezaka}, {Kiuchi}, {Kyutoku}, {Okawa},
  {Sekiguchi}, {Shibata} \& {Taniguchi}}{{Hotokezaka}
  et~al.}{2013}]{Hotokezaka2013}
{Hotokezaka} K.,  {Kiuchi} K.,  {Kyutoku} K.,  {Okawa} H.,  {Sekiguchi} Y.-i.,
  {Shibata} M.,    {Taniguchi} K.,  2013, \prd, 87, 024001

\bibitem[\protect\citeauthoryear{{Hut}, {Murphy} \& {Verbunt}}{{Hut}
  et~al.}{1991}]{hut91}
{Hut} P.,  {Murphy} B.~W.,    {Verbunt} F.,  1991, A\& A, 241, 137

\bibitem[\protect\citeauthoryear{{Jeans}}{{Jeans}}{1919}]{jeans1919}
{Jeans} J.~H.,  1919, \mnras, 79, 408

\bibitem[\protect\citeauthoryear{{Kinoshita} \& {Nakai}}{{Kinoshita} \&
  {Nakai}}{1999}]{kino1999}
{Kinoshita} H.,  {Nakai} H.,  1999, Celestial Mechanics and Dynamical
  Astronomy, 75, 125

\bibitem[\protect\citeauthoryear{{Knigge}, {Leigh} \& {Sills}}{{Knigge}
  et~al.}{2009}]{knigge09}
{Knigge} C.,  {Leigh} N.,    {Sills} A.,  2009, \nat, 457, 288

\bibitem[\protect\citeauthoryear{{Kozai}}{{Kozai}}{1962}]{koz62}
{Kozai} Y.,  1962, \aj, 67, 591

\bibitem[\protect\citeauthoryear{{Kraus}, {Ireland}, {Martinache} \&
  {Hillenbrand}}{{Kraus} et~al.}{2011}]{kraus11}
{Kraus} A.~L.,  {Ireland} M.~J.,  {Martinache} F.,    {Hillenbrand} L.~A.,
  2011, \apj, 731, 8

\bibitem[\protect\citeauthoryear{{Latham} \& {Milone}}{{Latham} \&
  {Milone}}{1996}]{latham96}
{Latham} D.~W.,  {Milone} A.~A.~E.,  1996, in {Milone} E.~F.,  {Mermilliod}
  J.-C.,  eds, The Origins, Evolution, and Destinies of Binary Stars in
  Clusters Vol.~90 of Astronomical Society of the Pacific Conference Series,
  {Spectroscopic Binaries Among the M67 Blue Stragglers}.
p.~385

\bibitem[\protect\citeauthoryear{{Lattimer} \& {Prakash}}{{Lattimer} \&
  {Prakash}}{2005}]{latt2005}
{Lattimer} J.~M.,  {Prakash} M.,  2005, Physical Review Letters, 94, 111101

\bibitem[\protect\citeauthoryear{{Lazzati}, {Perna}, {Morsony}, {Lopez-Camara},
  {Cantiello}, {Ciolfi}, {Giacomazzo} \& {Workman}}{{Lazzati}
  et~al.}{2018}]{Lazzati2018}
{Lazzati} D.,  {Perna} R.,  {Morsony} B.~J.,  {Lopez-Camara} D.,  {Cantiello}
  M.,  {Ciolfi} R.,  {Giacomazzo} B.,    {Workman} J.~C.,  2018, Physical
  Review Letters, 120, 241103

\bibitem[\protect\citeauthoryear{{Leigh} \& {Sills}}{{Leigh} \&
  {Sills}}{2011}]{leigh11}
{Leigh} N.,  {Sills} A.,  2011, \mnras, 410, 2370

\bibitem[\protect\citeauthoryear{{Leigh}, {Antonini}, {Stone}, {Shara} \&
  {Merritt}}{{Leigh} et~al.}{2016}]{leigh16}
{Leigh} N.~W.~C.,  {Antonini} F.,  {Stone} N.~C.,  {Shara} M.~M.,    {Merritt}
  D.,  2016, \mnras, 463, 1605

\bibitem[\protect\citeauthoryear{{Leigh} \& {Geller}}{{Leigh} \&
  {Geller}}{2013}]{leigh13}
{Leigh} N.~W.~C.,  {Geller} A.~M.,  2013, \mnras, 432, 2474

\bibitem[\protect\citeauthoryear{{Leigh}, {Geller}, {McKernan}, {Ford}, {Mac
  Low}, {Bellovary}, {Haiman}, {Lyra}, {Samsing}, {O'Dowd}, {Kocsis} \&
  {Endlich}}{{Leigh} et~al.}{2018}]{leigh18}
{Leigh} N.~W.~C.,  {Geller} A.~M.,  {McKernan} B.,  {Ford} K.~E.~S.,  {Mac Low}
  M.-M.,  {Bellovary} J.,  {Haiman} Z.,  {Lyra} W.,  {Samsing} J.,  {O'Dowd}
  M.,  {Kocsis} B.,    {Endlich} S.,  2018, \mnras, 474, 5672

\bibitem[\protect\citeauthoryear{{Lidov}}{{Lidov}}{1962}]{lid62}
{Lidov} M.~L.,  1962, \planss, 9, 719

\bibitem[\protect\citeauthoryear{{Liu} \& {Lai}}{{Liu} \&
  {Lai}}{2019}]{liu2019}
{Liu} B.,  {Lai} D.,  2019, \mnras, 483, 4060

\bibitem[\protect\citeauthoryear{{Lousto}, {Campanelli}, {Zlochower} \&
  {Nakano}}{{Lousto} et~al.}{2010}]{lou2010}
{Lousto} C.~O.,  {Campanelli} M.,  {Zlochower} Y.,    {Nakano} H.,  2010,
  Classical and Quantum Gravity, 27, 114006

\bibitem[\protect\citeauthoryear{{Mandel} \& {de Mink}}{{Mandel} \& {de
  Mink}}{2016}]{mand16}
{Mandel} I.,  {de Mink} S.~E.,  2016, \mnras, 458, 2634

\bibitem[\protect\citeauthoryear{{Marchant}, {Langer}, {Podsiadlowski},
  {Tauris} \& {Moriya}}{{Marchant} et~al.}{2016}]{march16}
{Marchant} P.,  {Langer} N.,  {Podsiadlowski} P.,  {Tauris} T.~M.,    {Moriya}
  T.~J.,  2016, A\& A, 588, A50

\bibitem[\protect\citeauthoryear{{Mardling} \& {Aarseth}}{{Mardling} \&
  {Aarseth}}{2001}]{mar01}
{Mardling} R.~A.,  {Aarseth} S.~J.,  2001, \mnras, 321, 398

\bibitem[\protect\citeauthoryear{{McKernan}, {Ford}, {Bellovary}, {Leigh}
  et~al.,}{{McKernan} et~al.}{2018}]{mckernan18}
{McKernan} B.,  {Ford} K.~E.~S.,  {Bellovary} J.,  {Leigh} N.~W.~C.,    et~al.,
  2018, \apj, 866, 66

\bibitem[\protect\citeauthoryear{{Mermilliod} \& {Mayor}}{{Mermilliod} \&
  {Mayor}}{1999}]{mermilliod99}
{Mermilliod} J.-C.,  {Mayor} M.,  1999, \aap, 352, 479

\bibitem[\protect\citeauthoryear{{Mermilliod}, {Rosvick}, {Duquennoy} \&
  {Mayor}}{{Mermilliod} et~al.}{1992}]{mermilliod92}
{Mermilliod} J.-C.,  {Rosvick} J.~M.,  {Duquennoy} A.,    {Mayor} M.,  1992,
  \aap, 265, 513

\bibitem[\protect\citeauthoryear{{Mikkola} \& {Merritt}}{{Mikkola} \&
  {Merritt}}{2006}]{mik06}
{Mikkola} S.,  {Merritt} D.,  2006, \mnras, 372, 219

\bibitem[\protect\citeauthoryear{{Mikkola} \& {Merritt}}{{Mikkola} \&
  {Merritt}}{2008}]{mik08}
{Mikkola} S.,  {Merritt} D.,  2008, \aj, 135, 2398

\bibitem[\protect\citeauthoryear{{Milone} et~al.,}{{Milone}
  et~al.}{2012}]{milone12}
{Milone} A.~P.,  et~al., 2012, A\& A, 540, A16

\bibitem[\protect\citeauthoryear{{Moe} \& {Kratter}}{{Moe} \&
  {Kratter}}{2018}]{moe18}
{Moe} M.,  {Kratter} K.~M.,  2018, \apj, 854, 44

\bibitem[\protect\citeauthoryear{{Naoz}}{{Naoz}}{2016}]{nao16}
{Naoz} S.,  2016, \araa, 54, 441

\bibitem[\protect\citeauthoryear{{Naoz} \& {Fabrycky}}{{Naoz} \&
  {Fabrycky}}{2014}]{naoz14}
{Naoz} S.,  {Fabrycky} D.~C.,  2014, \apj, 793, 137

\bibitem[\protect\citeauthoryear{{Naoz}, {Farr}, {Lithwick}, {Rasio} \&
  {Teyssandier}}{{Naoz} et~al.}{2013}]{naozf13a}
{Naoz} S.,  {Farr} W.~M.,  {Lithwick} Y.,  {Rasio} F.~A.,    {Teyssandier} J.,
  2013, \mnras, 431, 2155

\bibitem[\protect\citeauthoryear{{Naoz}, {Ghez}, {Hees}, {Do}, {Witzel} \&
  {Lu}}{{Naoz} et~al.}{2018}]{naoz2018}
{Naoz} S.,  {Ghez} A.~M.,  {Hees} A.,  {Do} T.,  {Witzel} G.,    {Lu} J.~R.,
  2018, \apj Lett, 853, L24

\bibitem[\protect\citeauthoryear{{Naoz}, {Kocsis}, {Loeb} \& {Yunes}}{{Naoz}
  et~al.}{2013}]{naoz2013}
{Naoz} S.,  {Kocsis} B.,  {Loeb} A.,    {Yunes} N.,  2013, \apj, 773, 187

\bibitem[\protect\citeauthoryear{{O'Leary}, {Kocsis} \& {Loeb}}{{O'Leary}
  et~al.}{2009}]{olea09}
{O'Leary} R.~M.,  {Kocsis} B.,    {Loeb} A.,  2009, \mnras, 395, 2127

\bibitem[\protect\citeauthoryear{{O'Leary}, {Meiron} \& {Kocsis}}{{O'Leary}
  et~al.}{2016}]{olea16}
{O'Leary} R.~M.,  {Meiron} Y.,    {Kocsis} B.,  2016, \apj Lett, 824, L12

\bibitem[\protect\citeauthoryear{{Ott}, {Eckart} \& {Genzel}}{{Ott}
  et~al.}{1999}]{ott1999}
{Ott} T.,  {Eckart} A.,    {Genzel} R.,  1999, \apj, 523, 248

\bibitem[\protect\citeauthoryear{{Patience}, {Ghez}, {Reid}, {Weinberger} \&
  {Matthews}}{{Patience} et~al.}{1998}]{patience98}
{Patience} J.,  {Ghez} A.~M.,  {Reid} I.~N.,  {Weinberger} A.~J.,    {Matthews}
  K.,  1998, \aj, 115, 1972

\bibitem[\protect\citeauthoryear{{Perets} \& {Fabrycky}}{{Perets} \&
  {Fabrycky}}{2009}]{perets09}
{Perets} H.~B.,  {Fabrycky} D.~C.,  2009, \apj, 697, 1048

\bibitem[\protect\citeauthoryear{{Perna}, {Lazzati} \& {Giacomazzo}}{{Perna}
  et~al.}{2016}]{Perna2016}
{Perna} R.,  {Lazzati} D.,    {Giacomazzo} B.,  2016, \apjl, 821, L18

\bibitem[\protect\citeauthoryear{{Perna}, {Wang}, {Farr}, {Leigh} \&
  {Cantiello}}{{Perna} et~al.}{2019}]{Perna2019}
{Perna} R.,  {Wang} Y.-H.,  {Farr} W.~M.,  {Leigh} N.,    {Cantiello} M.,
  2019, \apjl, 878, L1

\bibitem[\protect\citeauthoryear{{Petrovich} \& {Antonini}}{{Petrovich} \&
  {Antonini}}{2017}]{petr17}
{Petrovich} C.,  {Antonini} F.,  2017, \apj, 846, 146

\bibitem[\protect\citeauthoryear{{Pooley} \& {Hut}}{{Pooley} \&
  {Hut}}{2006}]{pooley06}
{Pooley} D.,  {Hut} P.,  2006, \apj Lett, 646, L143

\bibitem[\protect\citeauthoryear{{Prodan} \& {Murray}}{{Prodan} \&
  {Murray}}{2012}]{prodan12}
{Prodan} S.,  {Murray} N.,  2012, \apj, 747, 4

\bibitem[\protect\citeauthoryear{{Rafelski}, {Ghez}, {Hornstein}, {Lu} \&
  {Morris}}{{Rafelski} et~al.}{2007}]{rafelski2007}
{Rafelski} M.,  {Ghez} A.~M.,  {Hornstein} S.~D.,  {Lu} J.~R.,    {Morris} M.,
  2007, \apj, 659, 1241

\bibitem[\protect\citeauthoryear{{Raghavan} et~al.,}{{Raghavan}
  et~al.}{2010}]{ragh10}
{Raghavan} D.,  et~al., 2010, \apj Suppl, 190, 1

\bibitem[\protect\citeauthoryear{{Rasskazov} \& {Kocsis}}{{Rasskazov} \&
  {Kocsis}}{2019}]{rass2019}
{Rasskazov} A.,  {Kocsis} B.,  2019, arXiv e-prints, p. arXiv:1902.03242

\bibitem[\protect\citeauthoryear{{Rezzolla}, {Baiotti}, {Giacomazzo}, {Link} \&
  {Font}}{{Rezzolla} et~al.}{2010}]{Rezzolla2010}
{Rezzolla} L.,  {Baiotti} L.,  {Giacomazzo} B.,  {Link} D.,    {Font} J.~A.,
  2010, Classical and Quantum Gravity, 27, 114105

\bibitem[\protect\citeauthoryear{{Rodriguez}, {Amaro-Seoane}, {Chatterjee} \&
  {Rasio}}{{Rodriguez} et~al.}{2018}]{rod18}
{Rodriguez} C.~L.,  {Amaro-Seoane} P.,  {Chatterjee} S.,    {Rasio} F.~A.,
  2018, PRL, 120, 151101

\bibitem[\protect\citeauthoryear{{Samsing}}{{Samsing}}{2018}]{sam2018}
{Samsing} J.,  2018, \prd, 97, 103014

\bibitem[\protect\citeauthoryear{{Samsing}, {Askar} \& {Giersz}}{{Samsing}
  et~al.}{2018}]{samas18}
{Samsing} J.,  {Askar} A.,    {Giersz} M.,  2018, \apj, 855, 124

\bibitem[\protect\citeauthoryear{{Sana}}{{Sana}}{2017}]{sana2017}
{Sana} H.,  2017, in {Eldridge} J.~J.,  {Bray} J.~C.,  {McClelland} L.~A.~S.,
  {Xiao} L.,  eds, The Lives and Death-Throes of Massive Stars Vol.~329 of IAU
  Symposium, {The multiplicity of massive stars: a 2016 view}.
pp 110--117

\bibitem[\protect\citeauthoryear{{Sana} et~al.,}{{Sana}
  et~al.}{2013}]{sa2013AA}
{Sana} H.,  et~al., 2013, A\& A, 550, A107

\bibitem[\protect\citeauthoryear{{Sari}, {Piran} \& {Narayan}}{{Sari}
  et~al.}{1998}]{Sari1998}
{Sari} R.,  {Piran} T.,    {Narayan} R.,  1998, \apjl, 497, L17

\bibitem[\protect\citeauthoryear{{Secunda}, {Bellovary}, {Mac Low}, {Ford},
  {McKernan}, {Leigh} \& {Lyra}}{{Secunda} et~al.}{2018}]{secunda18}
{Secunda} A.,  {Bellovary} J.,  {Mac Low} M.-M.,  {Ford} K.~E.~S.,  {McKernan}
  B.,  {Leigh} N.,    {Lyra} W.,  2018, arXiv e-prints

\bibitem[\protect\citeauthoryear{{Shappee} \& {Thompson}}{{Shappee} \&
  {Thompson}}{2013}]{shapp2013}
{Shappee} B.~J.,  {Thompson} T.~A.,  2013, \apj, 766, 64

\bibitem[\protect\citeauthoryear{{Shibata}, {Duez}, {Liu}, {Shapiro} \&
  {Stephens}}{{Shibata} et~al.}{2006}]{Shibata2006}
{Shibata} M.,  {Duez} M.~D.,  {Liu} Y.~T.,  {Shapiro} S.~L.,    {Stephens}
  B.~C.,  2006, Physical Review Letters, 96, 031102

\bibitem[\protect\citeauthoryear{{Silsbee} \& {Tremaine}}{{Silsbee} \&
  {Tremaine}}{2017}]{sil17}
{Silsbee} K.,  {Tremaine} S.,  2017, \apj, 836, 39

\bibitem[\protect\citeauthoryear{{Sollima}}{{Sollima}}{2008}]{sollima08}
{Sollima} A.,  2008, \mnras, 388, 307

\bibitem[\protect\citeauthoryear{{Sollima}, {Beccari}, {Ferraro}, {Fusi Pecci}
  \& {Sarajedini}}{{Sollima} et~al.}{2007}]{sollima07}
{Sollima} A.,  {Beccari} G.,  {Ferraro} F.~R.,  {Fusi Pecci} F.,
  {Sarajedini} A.,  2007, \mnras, 380, 781

\bibitem[\protect\citeauthoryear{{The LIGO Scientific Collaboration} \& {the
  Virgo Collaboration}}{{The LIGO Scientific Collaboration} \& {the Virgo
  Collaboration}}{2018}]{ligo2018}
{The LIGO Scientific Collaboration} {the Virgo Collaboration} 2018, arXiv
  e-prints, p. arXiv:1811.12907

\bibitem[\protect\citeauthoryear{{Thompson}}{{Thompson}}{2011}]{thomp2011}
{Thompson} T.~A.,  2011, \apj, 741, 82

\bibitem[\protect\citeauthoryear{{Tokovinin}}{{Tokovinin}}{2014a}]{tok14a}
{Tokovinin} A.,  2014a, \aj, 147, 86

\bibitem[\protect\citeauthoryear{{Tokovinin}}{{Tokovinin}}{2014b}]{tok14b}
{Tokovinin} A.,  2014b, \aj, 147, 87

\bibitem[\protect\citeauthoryear{{Toonen}, {Perets} \& {Hamers}}{{Toonen}
  et~al.}{2018}]{toonen2018}
{Toonen} S.,  {Perets} H.~B.,    {Hamers} A.~S.,  2018, \aap, 610, A22

\bibitem[\protect\citeauthoryear{{Verbunt}, {van den Heuvel}, {van Paradijs} \&
  {Rappaport}}{{Verbunt} et~al.}{1987}]{verbunt87}
{Verbunt} F.,  {van den Heuvel} E.~P.~J.,  {van Paradijs} J.,    {Rappaport}
  S.~A.,  1987, Nature, 329, 312

\bibitem[\protect\citeauthoryear{{Wen}}{{Wen}}{2003}]{wen03}
{Wen} L.,  2003, \apj, 598, 419

\bibitem[\protect\citeauthoryear{{Zevin}, {Samsing}, {Rodriguez}, {Haster} \&
  {Ramirez-Ruiz}}{{Zevin} et~al.}{2018}]{zevin18}
{Zevin} M.,  {Samsing} J.,  {Rodriguez} C.,  {Haster} C.-J.,    {Ramirez-Ruiz}
  E.,  2018, arXiv:1810.00901

\end{thebibliography}

\end{document}